\documentclass[journal,letter,twoside]{IEEEtran}
\usepackage[utf8]{inputenc}
\usepackage[T1]{fontenc}

\usepackage{cite}
\usepackage{amsmath,amssymb,amsfonts}
\usepackage{bm}
\usepackage{algorithmic}
\usepackage{textcomp}
\usepackage{setspace}
\def\smallerspacecaption{}

\usepackage{floatrow}
\usepackage[caption=false,font=footnotesize]{subfig}

\usepackage[table, dvipsnames]{xcolor}  % Modern color package (used for text and tables)
\usepackage{fancyhdr}
\usepackage{graphicx}
\usepackage{url}
\usepackage{verbatim}
\usepackage{multirow}
\usepackage{hhline}
\usepackage[us]{datetime}
\usepackage{paralist}
\usepackage{color}
\usepackage{soul}
\usepackage[implicit=false,hidelinks]{hyperref}
\usepackage{stfloats}
\usepackage[ruled, vlined, norelsize]{algorithm2e} % The algorithms writing package
\SetAlFnt{\sf} % Set the algorithms font style to sans serif

\graphicspath{{figures/}}

\usepackage{soul}

\newdimen\arrayruleHwidth
\makeatletter
\def\Hline{\noalign{\ifnum0=`}\fi\hrule \@height \arrayruleHwidth
\futurelet \@tempa\@xhline}
\makeatother

\newcolumntype{P}[1]{>{\centering\arraybackslash}p{#1}}

\makeatletter
\def\blfootnote{\xdef\@thefnmark{}\@footnotetext}
\makeatother

\shortdate
\settimeformat{ampmtime}

\ifCLASSINFOpdf
\else
\fi
\begin{document}

\newcommand{\thetitle}{Spin-Orbit Torque Devices for Hardware Security: From Deterministic to Probabilistic Regime}

\title{\thetitle}

\author{Satwik~Patnaik$^*$,~\IEEEmembership{Student Member,~IEEE},
		Nikhil~Rangarajan$^*$,~\IEEEmembership{Student Member,~IEEE},
        Johann~Knechtel$^*$,~\IEEEmembership{Member,~IEEE}, 
        Ozgur~Sinanoglu,~\IEEEmembership{Senior Member,~IEEE},
        and~Shaloo~Rakheja,~\IEEEmembership{Member,~IEEE}
\thanks{$^*$S.\ Patnaik, N.\ Rangarajan, and J.\ Knechtel contributed equally.}        
\thanks{This work is an extension of~\cite{patnaik18_GSHE_DATE}.}
\thanks{S.\ Patnaik, N.\ Rangarajan, and S.\ Rakheja are with Department
of Electrical and Computer Engineering, Tandon School of Engineering, New York University, Brooklyn, NY, 11201, USA. \emph{Corresponding authors}: S.\
	Patnaik (sp4012@nyu.edu), N.\ Rangarajan (nikhil.rangarajan@nyu.edu), J.\ Knechtel (johann@nyu.edu), and S.\ Rakheja (shaloo.rakheja@nyu.edu).}% <-this % stops a space
\thanks{J.\ Knechtel and O.\ Sinanoglu (ozgursin@nyu.edu) are with Division of Engineering, New York University Abu Dhabi, Saadiyat Island, 129188, UAE.}% <-this % stops a space
}

\maketitle

\renewcommand{\headrulewidth}{0.0pt}
\thispagestyle{fancy}
\lhead{}
\rhead{}
\chead{\copyright~2019 IEEE.
This is the author's version of the work. It is posted here for personal use.
	Not for redistribution.	The definitive Version of Record is published in
IEEE Transactions on Computer-Aided Design of Integrated Circuits and Systems.}
\cfoot{}

\begin{abstract}
Protecting intellectual property (IP) has become a serious challenge for chip designers.
Most countermeasures are tailored for CMOS integration and tend to incur excessive 
overheads, resulting from additional circuitry or device-level modifications.
On the other hand,
power density is a critical concern for sub-50 nm nodes,
necessitating alternate design concepts.
Although initially tailored for error-tolerant applications,
imprecise computing has gained traction as a general-purpose design technique.
Emerging devices are currently being explored to implement ultra-low-power circuits for inexact computing applications.
In this paper, we quantify the security threats of imprecise computing using emerging devices. 
More specifically,
   we leverage the innate polymorphism and tunable stochastic behavior of spin-orbit torque (SOT) devices, particularly, the giant spin-Hall effect (GSHE) switch.
We enable IP protection (by means of logic locking and camouflaging) simultaneously for deterministic and probabilistic computing, directly at the GSHE 
device level.
We conduct a comprehensive 
security analysis using state-of-the-art Boolean satisfiability (SAT) attacks; this study demonstrates the superior resilience of our GSHE primitive when
tailored for deterministic computing.
We also demonstrate how
probabilistic computing can thwart most, if not all, existing SAT attacks.  Based on this finding, we propose an attack
scheme called probabilistic SAT (\emph{PSAT}) which can bypass the defense offered by logic locking and camouflaging for imprecise computing schemes.
Further, we illustrate how careful application of our GSHE primitive can remain secure even on the application of the PSAT attack.
Finally, we also discuss
side-channel attacks and 
invasive monitoring,
which are arguably even more concerning threats than SAT attacks.

\end{abstract}

\markboth{IEEE Transactions on Computer-Aided Design of Integrated Circuits and Systems}
{Patnaik \MakeLowercase{\textit{et al.}}: \thetitle}


\section{Introduction}
\label{sec:introduction}

\IEEEPARstart{T}{he} notion of imprecise computing already took root in 1956, with the seminal work
by von Neumann, where the concept of error was introduced as an essential part of any
computing system, subject to thermodynamical theory~\cite{von1956probabilistic}. Von Neumann further expounded on the control of errors in
simple automatons.
In 2007, the ITRS report~\cite{ITRS07} stated that ``relaxing the requirement of 100\% correctness
[...]
may
dramatically reduce costs of manufacturing, verification, and test. Such a paradigm shift is likely forced in any case by
technology scaling, which leads to more transient and permanent failures [...].''
Apart from the escalating strain on Moore's scaling, imprecise computing is further incentivized by the ``big-data
explosion'' and the rise of machine learning.
Both require intensive and large-scale computation, yet typically with some error tolerance, such that imprecise computing can
be leveraged~\cite{murphy12, gandomi15}.
At present, imprecise circuits are primarily tailored for error-tolerant applications including neural networks, voice recognition, and video processing.  
However, the proliferation and subsequent pervasiveness of imprecise computing is expected
in the near future, as
designers have already started to embrace computational errors as a means for achieving stringent requirements %on power dissipation  
in power- and data-intensive computational systems~\cite{sartori2011stochastic,bosio2017approximate}.
In this context, emerging devices including nanowire transistors, carbon-based electronics, spin-based computational elements,
offer further reduction in
power consumption as well as higher integration density compared to their CMOS counterparts~\cite{nikonov2013overview}.

\begin{figure}[tb]
\centering
  \includegraphics[scale=0.34]{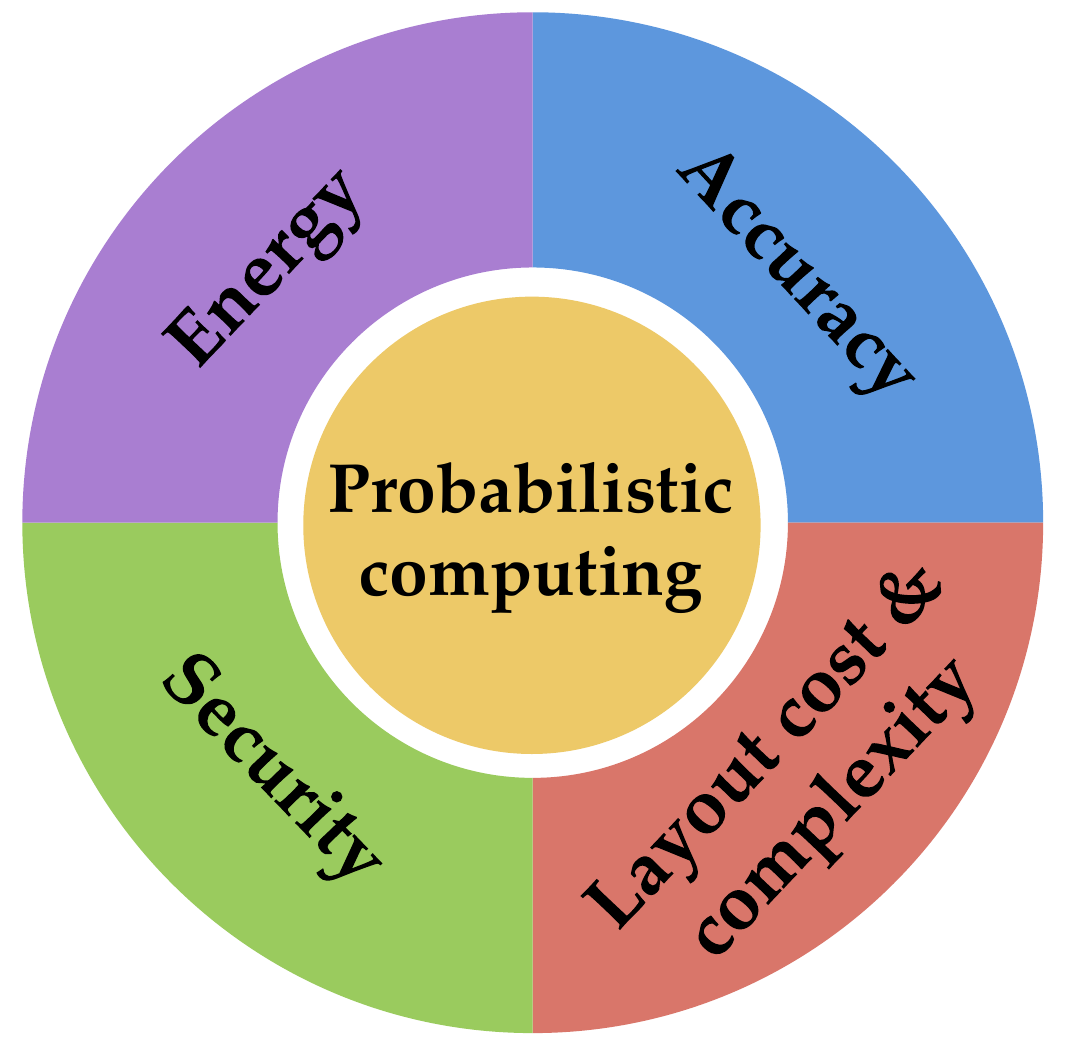}
  \caption{Interplay between accuracy, energy consumption, layout cost and complexity, and security for probabilistic
	  computing.
	  For any design, implementing larger parts using probabilistic logic helps to improve both the energy consumption and the resilience
	  against attack seeking to steal the underlying intellectual property.
	  However, this comes at the expense of reduced accuracy and
		  increased design complexity.
		  In any case, the accuracy of probabilistic gates can be tuned individually, which provides the designer with an additional
		  degree of freedom to exploit this interplay.}
  \label{fig:interplay}
  \smallerspacecaption
\end{figure}

Meanwhile, hardware security has become a major challenge due
to concerns such as theft of design intellectual property (IP), leakage of sensitive
data at runtime (via side channels or otherwise),
	counterfeiting of
chips, or insertion of hardware Trojans~\cite{rostami14}.
Recently it has been advocated that emerging devices can augment CMOS technology to advance hardware
security~\cite{ghosh2016spintronics,bi16_JETC, parveen17}, but very little focus has been given to imprecise computing so far.
Hence, it is crucial to discuss hardware security in the context of
imprecise computing systems, 
possibly built with emerging
devices.
Arguably the most promising aspect of 
many
emerging devices offered toward hardware security is their \emph{functional polymorphism}---a polymorphic gate can 
implement different Boolean functions,
	as determined by an
internal/external control mechanism~\cite{
		parveen17}.

In this work, we consider security as an essential design variable for emerging devices, and we examine the interplay between security,
   accuracy, layout cost and complexity, and energy (Fig.~\ref{fig:interplay}).
Although multi-functionality and polymorphism are inherent to spin-orbit torque (SOT) devices in general, we provision the giant spin-Hall
effect (GSHE) switch~\cite{datta2012non} here without loss of generality, since the technology of spin-Hall effect is more mature than other
charge to spin conversion devices currently~\cite{penumatcha2016impact, penumatcha2015spin}.
More specifically, we leverage the GSHE switch
demonstrated by Rangarajan \emph{et al.} for energy-efficient computing~\cite{rangarajan2017energy}.
We extend the scope of this GSHE switch toward hardware security, i.e.,
to build polymorphic gates for IP protection.
In doing so, we explore both the deterministic and the probabilistic regime.
While we choose the GSHE switch as a representative polymorphic spin gate for our work,
the concepts presented in this paper can be extended to other devices based on spin-orbit
coupling and the inverse Rashba-Edelstein effect, for instance the magnetoelectric spin orbit (MESO)
device~\cite{makarov16,manipatruni18}.

The structure and contributions of this paper can be summarized as follows.
\begin{enumerate}
\item We review imprecise computing, hardware security, and prior work in Sec.~\ref{sec:preliminaries}. We note that most prior works suffer
from low security resilience and/or high layout cost.
\item We design a polymorphic, GSHE-based
security primitive in detail in Sec.~\ref{sec:GSHE}.
The primitive provides strong security capabilities---given two inputs, all 16 possible Boolean functions can be packed within a single obfuscated
instance. Besides, the primitive can readily support deterministic and tunable probabilistic computing.
\item We elaborate on the protection provided by the primitive
against various attacks such as imaging-based reverse engineering,
side-channel attacks,
and analytical SAT attacks (Sec.~\ref{sec:security}).
Regarding SAT attacks, we conduct a comprehensive study (in the deterministic regime) to benchmark our primitive against prior defense schemes, which
are mainly based on magnetic devices.

\item The immunity of probabilistic computing against SAT attacks is explored in
Sec.~\ref{sec:security_prob}.
Most notably, here we present an advanced SAT attack, called PSAT,
which allows tackling IP protection for probabilistic circuits.
Using conventional SAT and PSAT, we reveal the trade-offs between accuracy, security, and energy for probabilistic computing.
Besides, we explore the resilience of polymorphic circuits.
\item In Sec.~\ref{sec:discussion}, among other aspects, we outline the prospects for
protecting industrial circuits using
	a hybrid CMOS-GSHE design style.
We anticipate that our proposed delay-aware protection can provide strong resilience against SAT attacks with negligible layout overheads.
\end{enumerate}

\section{Background and Motivation}
\label{sec:preliminaries}

\subsection{Imprecise Computing}
Three different branches of imprecise computing have emerged,
each leveraging unique techniques to harness noise and error for energy-efficient design: (1)~stochastic computing, (2)~approximate
computing, and (3)~probabilistic computing.

Stochastic computing is a paradigm that uses deterministic logic blocks for computation, but random binary bit streams. That is, the 
information is represented by statistical properties of the random bit stream implemented in space and
time~\cite{han2013approximate,alaghi18}. This way, for example, multiplication can be achieved using a single AND gate, albeit with
relatively low accuracy and long processing times. Digital and analog blocks for stochastic computing were introduced
in~\cite{gaines1969stochastic, alaghi2013survey}, and have since become popular for not only parallel, error-tolerant applications such as image-processing~\cite{alaghi2013stochastic},
	neuromorphic computing~\cite{gaba2013stochastic}, but also for general-purpose low-power designs~\cite{moons2014energy, sartori2011stochastic}.

Approximate computing is based on inexact logic for the least significant bits (LSBs) of any binary operation. This concept is implemented by altering the design at the circuit
level~\cite{han2013approximate}. Using techniques such as logic reduction or pass transistors with lower noise thresholds, approximate computing aims to forgo the accuracy of
LSBs to reduce power dissipation and circuit complexity~\cite{gupta2013low}. However, note that approximate computing does not
leverage the potential for non-determinism of the logic fabric itself.
Low-power
approximate full adder, constructed by transistor reduction in mirror-adder cells, and their application for signal processing were discussed in~\cite{gupta2011impact,
	gupta2013low}. A design and optimization framework for error and timing analysis of approximate circuits was proposed in~\cite{venkatesan2011macaco}. 
    Authors in~\cite{venkataramani2014axnn} design an energy-efficient approximate neural network by selective replacement of least significant neurons in the network with imprecise versions. 

Probabilistic computing relies on noisy gates that exploit the thermal randomness present in any computing system and, hence, implement an
inherently stochastic behavior~\cite{han2013approximate}.
Due to the stochastic behavior, key design steps such as verification are naturally more challenging than they are for deterministic
computing~\cite{lee18}.
Probabilistic computing with
CMOS logic is realized by using a noise source (often external), which introduces metastability and error in the binary CMOS switches~\cite{cheemalavagu2005probabilistic}. 
Spintronic gates, such as the GSHE gate~\cite{datta2012non} and the all-spin logic (ASL) gate~\cite{behin2010proposal}, are intrinsically random, without the
need for external noise sources, due to their stochastic magnetization dynamics~\cite{brown1963thermal, d2006midpoint}. A detailed study of
probabilistic GSHE logic is presented in~\cite{rangarajan2017energy}, which quantifies the energy savings for various operating accuracies
and also outlines error models for complex logic gates.

\subsection{Simple Case Study for Approximate Computing}
\label{sec:case_study_app}

Here, we present a simple case study illustrating 
the power-accuracy trade-off in the GSHE device~\cite{rangarajan2017energy}. 
We follow the fundamental rule of thumb for low-power approximate computing, i.e., the most significant bits (MSBs) shall have a higher
priority than the 
LSBs while trading off computational accuracy for power savings.
By assigning probabilistic behavior to the logic gates which only affect the LSBs, we restrict the loss in computational accuracy.

Consider a 32-bit adder
synthesized using GSHE gates
that initially operate deterministically.
Now, the logic gates that impact the
computation of the LSBs---the lower 10 bits in this example---can be made to behave probabilistically by operating them at sub-critical input
currents~\cite{rangarajan2017energy}. Considering an error rate of
10\% for those gates,
    the worst-case error for the 32-bit addition is only about 0.000024\%, whereas
the power savings per gate are 50\%, and in total
    about 5\%.
The GSHE gate in this work
can be tuned to incur a power consumption
of 0.2125 $\mu$W for the deterministic regime
(Sec.~\ref{sec:characterization}),
whereas the same gate when operating with an output
accuracy of 90\%, consumes only 0.1071 $\mu$W~\cite{rangarajan2017energy}.

Note that this discussion covers only
accuracy and power, and the aspect of security will be introduced later.
\subsection{Hardware Security}

With the advent of globalization affecting the supply chain of integrated circuits (ICs),
hardware security has emerged as a critical concern.
The exposure to potential adversaries in any of the third parties tasked for design, manufacturing, and testing has
escalated~\cite{rostami14}. These adversaries may seek to
(i) reverse engineer (RE) the ICs, (ii) counterfeit the ICs, (iii) steal the underlying intellectual
property (IP), or (iv) insert some hardware Trojans.
Regarding the IP-centric threats, it has been estimated that
billions of dollars in revenue are lost every year~\cite{karri2017physical}.
To mitigate these threats,
	  several schemes have been proposed; they are mostly based either on
camouflaging, logic locking, or split manufacturing.

\begin{figure}[tb]
\centering
  \includegraphics[scale=0.12]{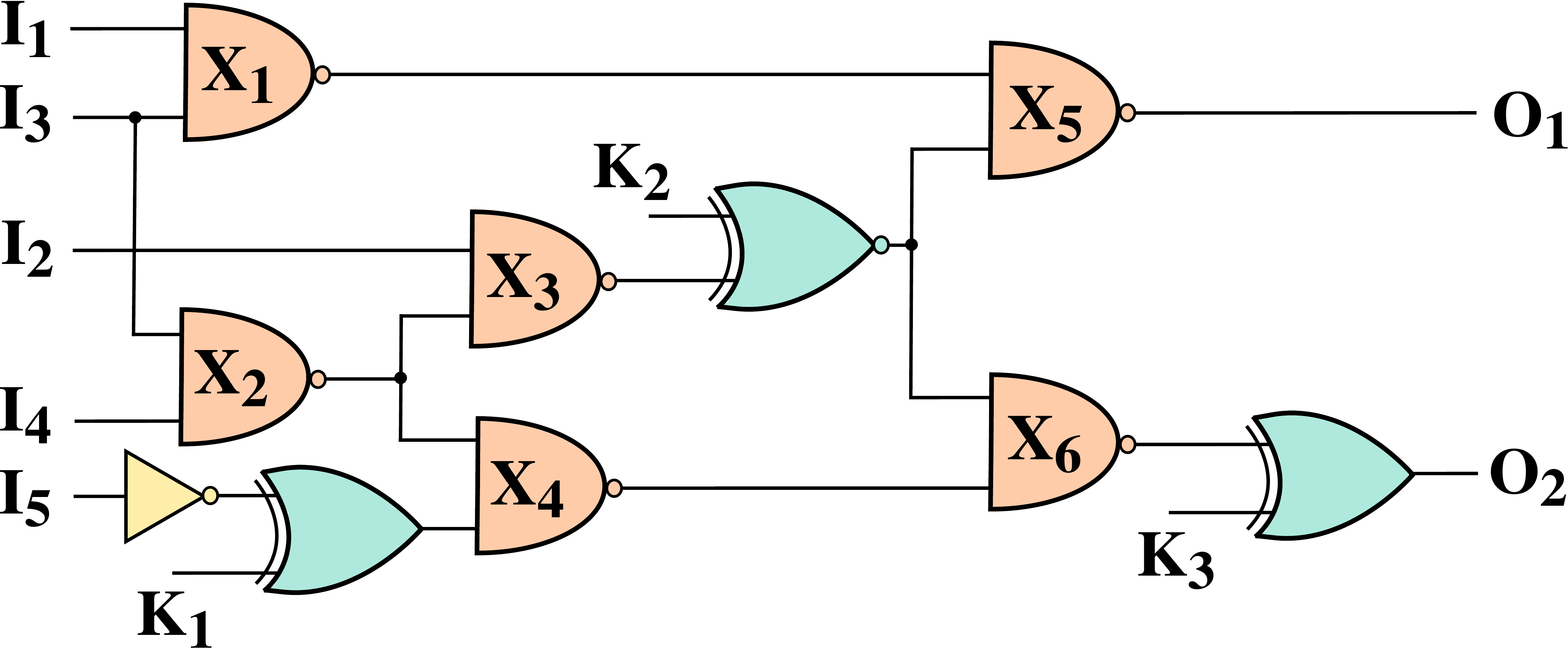}
  \caption{The \emph{ISCAS-85} benchmark \emph{c17} with three key gates: $\text{K}_1$, $\text{K}_2$, and $\text{K}_3$.}
  \label{fig:c17_example}
  \smallerspacecaption
\end{figure}

Camouflaging seeks
to mitigate RE attacks; wherein the layout-level appearance of the IC is altered in  a manner such that it becomes intractable to
decipher its underlying functionality and IP~\cite{vijayakumar16}.
For CMOS integration, various techniques have been proposed, e.g., look-alike gates~\cite{rajendran13_camouflage},
threshold-voltage-dependent (TVD) camouflaging~\cite{nirmala16, erbagci16}, and camouflaging of the
interconnects~\cite{patnaik17_Camo_BEOL_ICCAD}.
Emerging devices have been recently considered for camouflaging as well, e.g.,	 in~\cite{zhang2015giant,alasad2017leveraging,winograd2016hybrid,yang18}; see also Sec.~\ref{sec:prior_art} and
	 Sec.~\ref{sec:characterization}.

Logic locking (also known as logic encryption)
obfuscates the IP functionality rather than the device-level 
layout~\cite{roy10, yasin16_SARLock, xie16_SAT}.
Here, the designer obfuscates
the netlist by inserting additional key gates such that the original functionality can only be restored once the correct key bits
are applied. The key bits are 
programmed
into a tamper-proof memory after fabrication; this is to hinder attacks during manufacturing time as well
as in the field.
However, realizing tamper-proof memories is a practical challenge~\cite{tuyls06,anceau17}; emerging spin devices can be promising
here as well~\cite{ghosh2016spintronics}.

Analytical attacks targeting camouflaged (or locked) ICs were initially introduced in~\cite{subramanyan15, massad15}.  Most analytical
attacks are based on some notion of Boolean satisfiability (SAT) where a relatively small set of discriminating input patterns (DIPs) may
suffice to resolve the functionality of camouflaged gates (or locking keys); see also Sec.~\ref{sec:SAT_attack_background}.
Several SAT-attack resilient techniques were recently proposed, e.g.,~\cite{yasin16_SARLock, xie16_SAT,
	li16_camouflaging}. These works seek to impose exponential computation complexity for the SAT solver.
Still, most of these techniques are 
vulnerable to some degree when subject to tailored attacks such as~\cite{shamsi18_TIFS, shen17, xu17_BDD}.
Besides, we note that prior art on SAT attacks tender exclusively to deterministic computing and circuits.
In this paper, among other contributions, we extend the
scope of SAT attacks to probabilistic circuits.

Physical attacks range from non-invasive (e.g., power side-channel attacks) and semi-invasive (e.g., localized fault-injection attacks) to invasive attacks (e.g., RE,
		microprobing the frontside/backside)~\cite{wang17_probing}.
While such attacks require more sophisticated tools and know-how than analytical attacks,
their potential is widely acknowledged to be more severe.
Such attacks are also promising for extracting sensitive data at runtime, even from secured chips, e.g.,~\cite{skorobogatov12,courbon16}.

\subsection{Boolean Satisfiability and Related Attacks}
\label{sec:SAT_attack_background}

The problem of Boolean satisfiability (SAT) is an NP-complete problem that determines if a given propositional Boolean formula, usually
expressed in
its conjunctive normal form (CNF), can be satisfied by any combination of values assumed by the variables of the
formula~\cite{weissenbacher2014boolean}.
In case one or more such
combinations result in a ``true'' evaluation of the Boolean formula, then it is termed satisfiable and otherwise, unsatisfiable.
A SAT solver is based on an algorithm which heuristically sweeps through the solution space of the Boolean formula to check if any
particular combination of variable assignments can satisfy the formula. Such SAT solvers have become quite prevalent for 
combinatorial optimization, software verification, and cryptanalysis
applications~\cite{moskewicz2001chaff, mironov2006applications}.
More recently, SAT solvers have also been tailored for the field of hardware security, namely to resolve/attack logic-locked or camouflaged
circuits~\cite{subramanyan15, massad15, yu17, shamsi18_TIFS, wang18}.
Such SAT attacks are successful once the attacker can resolve the key bits of the locking scheme or the functionality of all the camouflaged
gates.  Note that the various possible, obfuscated functionalities for camouflaging
can be modeled as key bits as well.
In other words, 
camouflaging and logic locking are interchangeable in terms of analytical modeling~\cite{yu17, yasin15_IDT}.

\begin{table*}[tb]
\scriptsize
\caption{SAT Attack on the Benchmark \emph{c17}, Locked as in Fig.~\ref{fig:c17_example}
\label{tab:SAT_attack_example}
}
\begin{center}
\smallerspacecaption
\begin{tabular}{ P{1.5cm} | P{1cm} | P{0.5cm} | P{0.5cm} | P{0.5cm} | P{0.5cm} | P{0.5cm} | P{0.5cm} | P{0.5cm} | P{0.5cm} | P{6.6cm} }
\hline
{\textbf{Input Patterns}} & \textbf{Oracle Output}& \multicolumn{8}{c|}{{\multirow{2}{*}{\textbf{Output for Different Key Combinations}}}}
&{\multirow{2}{*}{\textbf{Inference}}}\\
\cline{2-11}
$\bm{\text{\textbf{I}}_1\text{\textbf{I}}_2\text{\textbf{I}}_3\text{\textbf{I}}_4\text{\textbf{I}}_5}$&$\bm{\text{\textbf{O}}_1\text{\textbf{O}}_2}$ &$\bm{k_0}$ &$\bm{k_1}$ &$\bm{k_2}$ &$\bm{k_3}$ &$\bm{k_4}$ &$\bm{k_5}$ &$\bm{k_6}$ &$\bm{k_7}$ &\\
\hline
\hline
$00000$ &$00$  &$01$ &$00$ &$10$ &$11$ &$00$ &$01$ &$10$ &$11$ & \\
\hline
$00001$ &$01$  &$00$ &$01$ &$10$ &$11$ &$01$ &$00$ &$10$ &$11$ & \\
\hline
$00010$ &$11$  &$10$ &$11$ &$01$ &$00$ &$10$ &$11$ &$00$ &$01$ & \\
 \hline
$00011$ &$11$  & $10$&$11$ &$00$ &$01$ &$10$ &$11$ &$01$ &$00$ & \\
\hline
$00100$ &$00$ &\cellcolor{red!25}$01$ &$00$ &\cellcolor{red!25}$10$ &\cellcolor{red!25}$11$ &$00$ &\cellcolor{red!25}$01$
&\cellcolor{red!25}$10$ &\cellcolor{red!25}$11$ &Iteration 1: $k_0, k_2, k_3, k_5, k_6, k_7$ are pruned \\
\hline
$00101$ &$01$  &$00$ &$01$ &$10$ &$11$ &$01$ &$00$ &$10$ &$11$ & \\
\hline
$00110$ &$11$  &$10$ &$11$ &$01$ &$00$ &$10$ &$11$ &$00$ &$01$ & \\
\hline
$00111$ &$11$ &$10$ &\cellcolor{green!25}$11$ &$00$ &$01$ &\cellcolor{red!25}$10$ &$11$ &$01$ &$00$ &Iteration 2: $k_4$ is pruned
$\Rightarrow k_1$ is inferred as correct key \\
\hline
$\dots$ &$\dots$ &$\dots$ &$\dots$ &$\dots$ &$\dots$ &$\dots$ &$\dots$ &$\dots$ &$\dots$ & \\
\hline
$11111$ &$10$  &$11$ &$10$ &$10$ &$11$ &$11$ &$10$ &$10$ &$11$ & \\
\hline
\end{tabular}

\\[1mm]
	Labels $k_0$--$k_7$ represent all possible combinations of key bits, from
	$000$ to $111$, and columns denote the corresponding outputs.
	$k_1$ is the correct key.
\end{center}
\smallerspacecaption
\end{table*}

Next, we provide a simple example that illustrates how SAT attacks decipher a locked netlist in general, namely by repeated
iterations over the key space. Consider the benchmark circuit \emph{c17} shown in Fig.~\ref{fig:c17_example}, which has been locked with three
key bits: $\text{K}_1$, $\text{K}_2$, and $\text{K}_3$.
Now, the attack procedure is to stepwise fix a particular input pattern and then iterate over various possible keys, eliminating those whose variable
assignments cannot satisfy the Boolean formula.
For example, in Table~\ref{tab:SAT_attack_example}, the input pattern ``00100'' is chosen (either randomly or heuristically) in the first
iteration.
The corresponding output is ``00'', as obtained from an oracle (i.e., a working chip queried with the input).
Accordingly, some keys cannot result in satisfiable assignments, namely
$k_0, k_2, k_3, k_5, k_6$, and $k_7$. These keys are pruned, and in the
same way, the second iteration prunes key $k_4$. This leaves $k_1$ as the last remaining key, which is returned as the attack solution.

It is important
to note that the outlined attack flow remains purposefully generic and abstract. Actual SAT attacks all apply various heuristics and techniques to
efficiently tackle the solution space, avoiding brute-force behavior. Interested readers are referred
to~\cite{subramanyan15, massad15, shamsi18_TIFS, yu17, wang18}.

\subsection{Prior Art and Limitations}
\label{sec:prior_art}

Now, we briefly review some prior art and their limitations. A more detailed comparison in terms of power and delay, and
(lack of) resilience against SAT attacks are provided in Sec.~\ref{sec:characterization} and Sec.~\ref{sec:det_SAT_study}
(Table~\ref{tab:SAT_pramod}), respectively.

In~\cite{zhang2015giant}, Zhang \emph{et al.}
implemented a low-power and versatile gate
using a GSHE-based magnetic tunnel junction (MTJ) as the basic switching element.
However, this device is not explicitly tailored for security;
it is unable to support logic locking by itself, as it is
not polymorphic.
More concerning
is the limitation to only four possible Boolean functions, which renders this primitive weak against SAT attacks.

Alasad~\emph{et al.}~\cite{alasad2017leveraging} use ASL to design 
three different security primitives, supporting three sets of camouflaged functionalities: INV/BUF, XOR/XNOR, and AND/NAND/OR/NOR.
The layouts of the three primitives are unique; they can be readily distinguished
by imaging-based RE tools, which also eases subsequent SAT attacks.
Besides, the primitives suffer from relatively high power consumption of $\sim 350$ $\mu$W at ns delays.

Winograd~\emph{et al.}~\cite{winograd2016hybrid} introduced a spin-transfer torque (STT)-based reconfigurable lookup table (LUT), explicitly addressing hardware security.
However, their approach
	falls short regarding resilience against SAT attack.
Since the authors did not report on any SAT attack themselves, we conducted
exploratory experiments ourselves. For example, we protect the \emph{ITC-99} benchmark \emph{s38584} 
according to their scheme
and observe that the resulting layouts can be decamouflaged in less than 30 seconds (i.e., average SAT runtime over 100 runs
		of random gate selection according to~\cite{winograd2016hybrid}).
This weak resilience stems from the limited use of their STT-LUT primitive
to curb power, performance, and area (PPA) overheads.

Yang~\emph{et al.}~\cite{yang18} recently proposed an SOT-based design for reconfigurable LUTs.
Their concept is tailored for obfuscation; 
it can support all 16 possible Boolean functions for two inputs, like ours.
However, the authors neglect powerful SAT attacks. Based on overly optimistic assumptions regarding the
attacker's capabilities (presumably to curb PPA cost as well), the authors limit their study to the obfuscation of 16/32/64 gates. In
experiments similar to those we conducted for~\cite{winograd2016hybrid}, we found that such small-scale obfuscation is easily resolved.

As for CMOS-centric camouflaging,
most schemes are static (i.e., not polymorphic) and tend to
incur a high layout cost. For example,
the static look-alike NAND-NOR-XOR gate proposed by Rajendran~\emph{et al.}~\cite{rajendran13_camouflage} induces overheads of 4$\times$ in area, 5.5$\times$ in power, and 1.6$\times$ in delay (compared to a
		regular two-input NAND gate). The TVD full-chip camouflaging as proposed in~\cite{erbagci16} still induces overheads of 14\%, 82\%, and 150\% in PPA, respectively.
As a result, such schemes are limited to a cost-constrained and selective application, which has severe implications for security
(Sec.~\ref{sec:security}).

Finally, we acknowledge that Koteshwara \emph{et al.} recently proposed dynamic obfuscation at the system level~\cite{koteshwara17},
albeit their work focuses only on CMOS implementation.
We anticipate that
inherently polymorphic gates, such as the GSHE device, can advance such a scheme.
Furthermore, McDonald \emph{et al.}~\cite{mcdonald16} advocate runtime polymorphism for IP protection, albeit without
any security analysis using SAT attacks and without any implementation details toward polymorphic devices.

\section{Design of A Giant Spin Hall Effect (GSHE) Security Primitive}
\label{sec:GSHE}

Protection schemes based on emerging devices can be competitive, even when compared to regular, unprotected CMOS circuits.
Leveraging GSHE is one approach among many to realize SOT-based magnetic devices. The GSHE switch has been studied for a while,
and its understanding is relatively mature.
The SOT phenomena has been experimentally measured in several magnetic and non-magnetic bilayers at 300K~\cite{miron2010current,
fukami2016magnetization}. Likewise, the read-out mechanism in the GSHE device
has been experimentally demonstrated in similar magnetic structures~\cite{almasi2015enhanced, li2014electric}.
Experiments are currently underway to integrate the read and write circuitry in the GSHE device to realize Boolean logic~\cite{penumatcha2016impact,penumatcha2015spin}.

Note that truly polymorphic gates such as the GSHE switch
can inherently support both
camouflaging and locking due to the following reasons.
First, owing to their uniform device-level layout, the actual function of a polymorphic gate is hard to determine from its physical implementation,
	particularly when optical-imaging-based RE techniques are used.
	Second, the actual function
	is dependent on control currents and voltages, which can act as key inputs.
Hence, the notions of locking and camouflaging are
used interchangeably
in the remainder of this work.

\subsection{Structure and Operating Principle of the GSHE Switch}
\label{sec:GSHE_operation}

\begin{figure}[tb]
\centering
  \includegraphics[width=.85\textwidth]{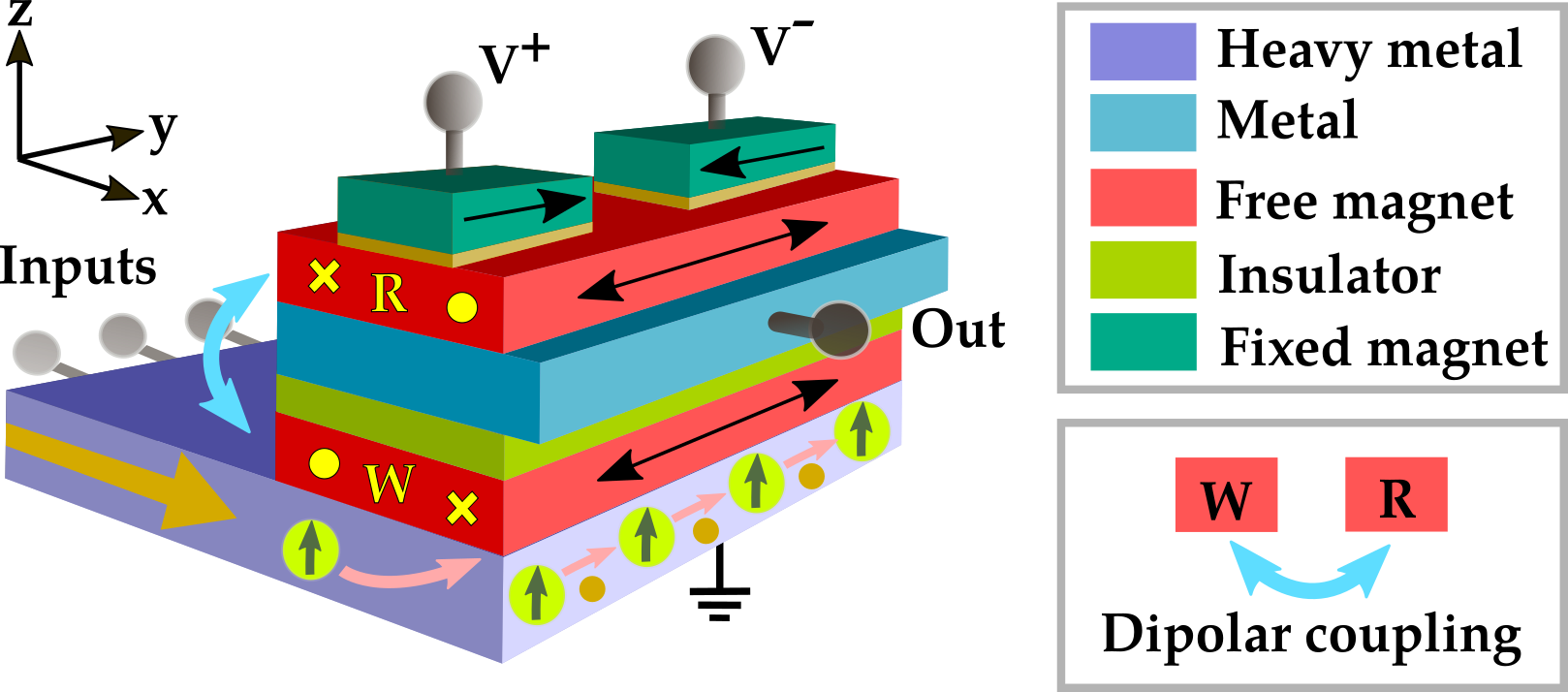}
  \caption{Structure of the GSHE switch. The concept is derived from~\cite{rangarajan2017energy}, but here we adopt a stacked integration to maximize the dipolar coupling.
  \label{fig:GSHE_gate}
  }
  \smallerspacecaption
\end{figure}

The GSHE switch, which is at the heart of the proposed primitive, is constructed by combining a heavy-metal spin-Hall layer, such as tantalum, tungsten, platinum or palladium, with a magnetic tunnel junction (MTJ) arrangement (Fig.~\ref{fig:GSHE_gate}).
Above the heavy-metal layer are two nanomagnets for write and read modes (W-NM and R-NM, red).
The W-NM is separated from the output terminal via an insulating oxide layer (green).
On top of the R-NM sit two fixed ferromagnetic layers (dark green) with anti-parallel magnetization
directions.

The switch relies on the spin-Hall effect~\cite{hirsch1999spin} for generating and amplifying the spin current input, and the magnetic dipolar
coupling phenomenon~\cite{kani2016model} to magnetically couple R-NM and W-NM, while keeping them electrically isolated.
Thereby, the R-NM and W-NM are coupled.
A charge current through
the bottom heavy-metal layer (along $\hat{x}$) induces a spin current in the transverse direction (along $\hat{y}$), which is used to switch the magnetization state of
the W-NM. The dipolar coupling field then causes the R-NM to switch its orientation. That is because in the
	presence of magnetic dipolar coupling, the minimum energy state is the one in which the W-NM and the R-NM
are anti-parallel to each other~\cite{datta2012non}.
The final magnetization state of the R-NM is read off using a differential MTJ setup. The logic (1 or 0) is encoded in the direction
of the electrical output current (+I or -I). The current direction depends on the relative orientations of the fixed magnets in the MTJ stack with respect to the final magnetization of the R-NM.
The parallel path offers a lower resistance for a charge current passing either from the MTJ contact to the output terminal
or vice versa (i.e., from the output terminal to the MTJ contact).
Hence, depending on the polarity of the read-out voltage
applied to the low-resistance path
(i.e., either V$^+$ or V$^-$),
the output current either flows inward or outward,
	representing the logic encoding of the GSHE
switch operation.

\begin{figure}[tb]
    \centering
    \includegraphics[width=\textwidth]{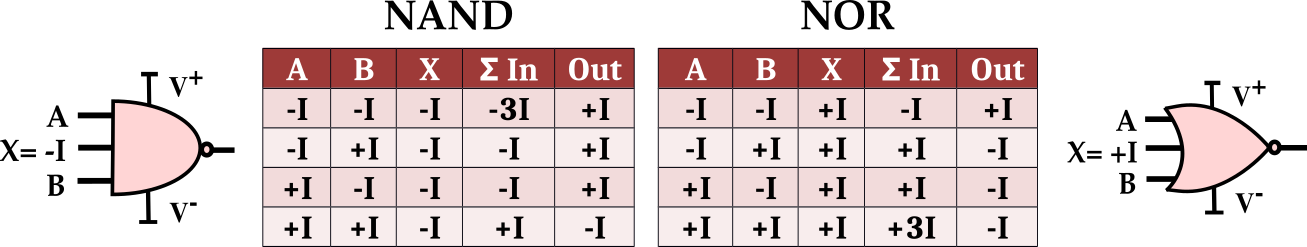}
    \caption{The current-centric truth tables for NAND and NOR functionalities, with inputs A and B (X is a control signal). As always the case for our GSHE-based primitive, logic 1/0 is represented by an output current +I/-I.
    \label{fig:GSHE_NAND_NOR}
    }
  \smallerspacecaption
\end{figure}

This basic GSHE device can readily implement a BUF or INV gate (buffer or inverter operations). To
realize more complex multi-input logic gates, 
a tie-breaking control signal $X$ with a fixed amplitude and polarity is applied in addition to the primary input signals at the input terminal.
That is, the input of the GSHE switch (or, more generally, any SOT-driven magnetic switch) is additive in nature.
The polarities of the control signal and the MTJ voltage polarities
are used to permute between different Boolean operations (Fig.~\ref{fig:GSHE_NAND_NOR}).
See also Sec.~\ref{sec:primitive} and its Fig.~\ref{fig:GSHE_gates} for all 16 possible Boolean gates.

The GSHE switch is a noisy polymorphic device,
whose probability for output correctness depends on the input spin current's amplitude and duration, i.e., the outputs generated by the
	previous logic stages.
In general, the control signal $X$ is used to set the functionality for any current stage ($n^{th}$ stage), and
	the output correctness probability for each next stage ($(n+1)^{th}$ stage) is as follows~\cite{rangarajan2017energy}:
\begin{equation}
 \mathcal{P}_{\text{correct}}^{\text{n+1}} = \mathcal{P}_{\text{flip}}\left(I_{\text{sX}}+\Sigma_{\text{i}}^{N_{\text{input}}}\frac{\beta \Delta G_{\text{i}}^{\text{n}} V_{\text{i}}^{\text{n}}}{1+rG_{\text{i}}^{\text{n}}}\right)\left[f\right] + 1.\left[1-f\right]
    \label{eq:Pcorr2}
\end{equation}
\noindent where $\mathcal{P}_{\text{flip}}$ is the probability for flipping of the W-NM in the $(n+1)^{\text{th}}$ stage's GSHE switch.
This probability itself is a function of the current supplied by the $n^{\text{th}}$ stage and the magnitude of the control spin current $I_{\text{sX}}$.
The relationship between the output current of a particular stage and the voltage supply in the MTJ
arrangement of that stage is according to~\cite{rangarajan2017energy}, wherein $\beta$ is the spin-Hall current amplification factor, $G^{\text{n}}$
is the MTJ conductance for the $n^{\text{th}}$
stage, $V^{\text{n}}$ is the MTJ voltage for the $n^{\text{th}}$ stage, and $r$ is the resistance of the spin Hall layers.
The function $f$ represents the Boolean function to be implemented, and
establishes the condition for flipping of the magnetization
state.

\subsection{Characterization and Comparison of the GSHE Switch}
\label{sec:characterization}

\begin{figure}[tb]
\centering
\includegraphics[width=.72\textwidth]{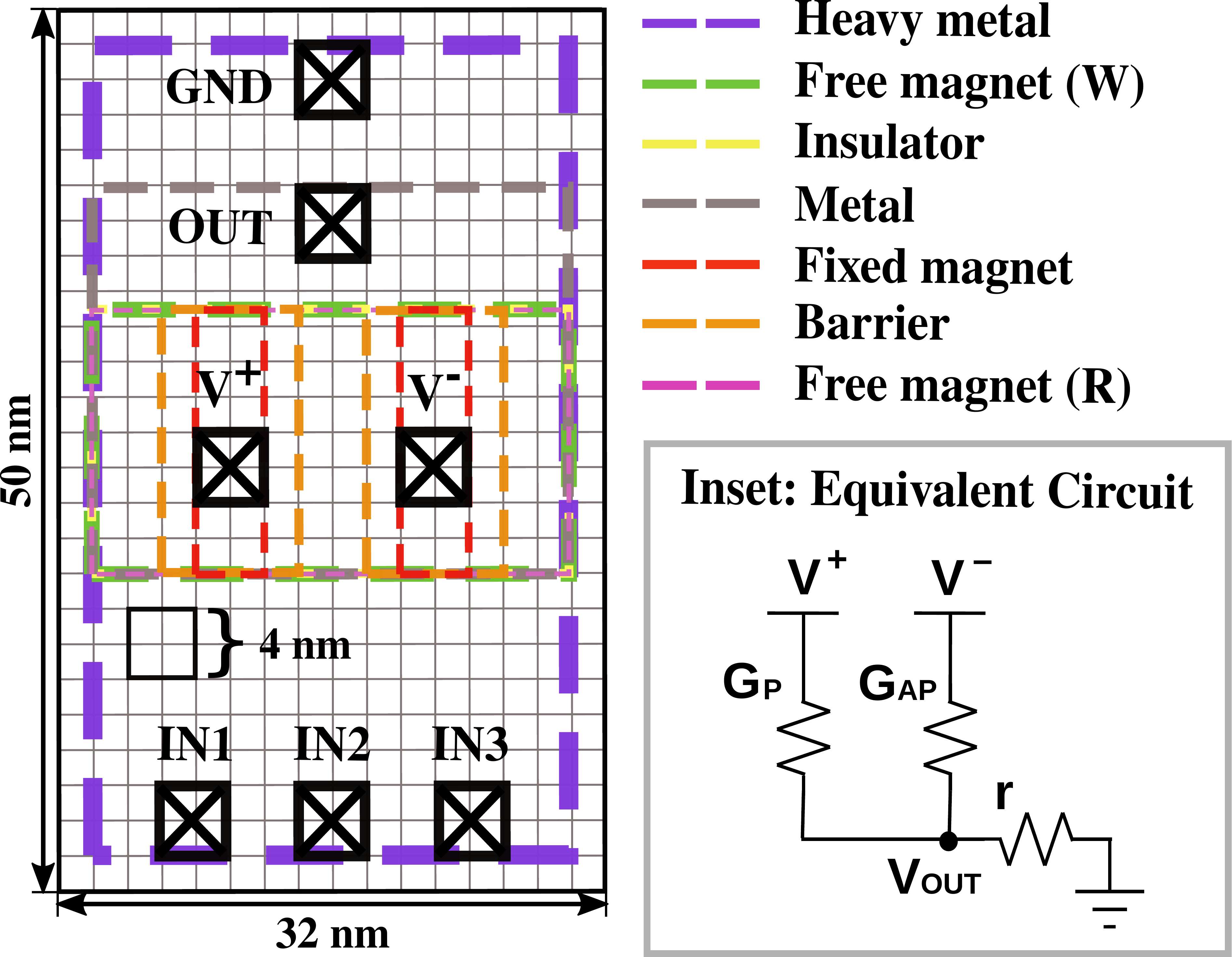}
\caption{The 
layout of the GSHE switch constructed according to the design rules for beyond-CMOS devices formulated in~\cite{nikonov2013overview}. Inset shows the equivalent circuit of the GSHE switch, derived from~\cite{datta2012non}.
	The power dissipation in the
		equivalent circuit
		is dictated by
the resistance $r$ of the heavy metal as well as the conductances of the anti-parallel, high-resistance path ($G_{AP}$) and the parallel, low-resistance path ($G_P$) composed of the fixed
ferromagnets.}
\label{fig:Layout_GSHE}
  \smallerspacecaption
\end{figure}

The layout of the GSHE switch 
(Fig.~\ref{fig:Layout_GSHE}) is drawn based on the design rules for beyond-CMOS devices~\cite{nikonov2013overview}, i.e., in units of maximum misalignment length
$\lambda$. The
area of the GSHE switch is accordingly estimated to be $0.0016 \mu$m$^2$.

The material parameters for the GSHE switch considered are given in Table~\ref{tab:GSHE_parameters}.
A
spin current ($I_{S}$) of at least 20$\mu$A is required for deterministic computing,
as compared to the sub-critical currents sufficient for probabilistic computing~\cite{rangarajan2017energy}.

\begin{table}[tb]
\scriptsize
\renewcommand{\arraystretch}{1.15}
\caption{Material Parameters of the GSHE Switch
}
\setlength{\tabcolsep}{1.4mm}
\begin{center}
\smallerspacecaption
\begin{tabular}{c|c}
\hline
{\textbf{Parameter}} & {\textbf{Value}}  \\
\hline
\hline
Volume of nanomagnets (NM) & ($28\times 15\times 2$) nm$^3$~\cite{rangarajan2017energy} \\
\hline
\multirow{2}{*}{Saturation magnetization $M_{\text{S}}$ of NM } &  $10^6$ A/m (W-NM)~\cite{rangarajan2017energy}\\
	   & $5\times 10^5$ A/m (R-NM)~\cite{rangarajan2017energy}\\
\hline
\multirow{2}{*}{Uniaxial energy density $K_{\text{u}}$ of NM} & $2.5\times 10^4$ J/m$^3$ (W-NM)~\cite{rangarajan2017energy} \\
& $5\times 10^3$ J/m$^3$ (R-NM)~\cite{rangarajan2017energy} \\
 \hline
Spin current $I_{\text{S}}$, determ.\ switching
& 20 $\mu$A~\cite{rangarajan2017energy}\\
\hline
Resistance area product $\text{RAP}$  &$1\>\> \Omega \mu$m$^2$ \cite{maehara2011tunnel}\\
\hline
Tunneling magnetoresistance $\text{TMR}$ &$170\%$ \cite{maehara2011tunnel}\\
\hline
Parallel conductance $G_{\text{P}}$ & $420\>\>\mu$S\\
\hline
Anti-parallel conductance $G_{\text{AP}}$ & $155.6\>\>\mu$S\\
\hline
Resistivity of heavy metal (HM) $\rho$ &$5.6\times 10^{-7} \Omega$--m\\
\hline
Spin-Hall angle $\theta_{\text{SH}}$ of HM &$0.4$~\cite{hao2015giant}\\
\hline
Thickness $t_{\text{HM}}$ of HM & $1$ nm\\
\hline
Internal gain $\beta$ of HM & $0.4\times(15\;$nm$/1\;$nm$)$\\
$\beta = \theta_{\text{SH}}\times(w_{\text{NM}}/t_{\text{HM}})$ & $=6$ \\
\hline
Resistance $r$ of HM & $\approx 1\>\> k\Omega$ \\
\hline
\end{tabular}

\label{tab:GSHE_parameters}
\end{center}
\smallerspacecaption
\smallerspacecaption
\end{table}

\begin{figure}[tb]
\centering
\includegraphics[width=.85\textwidth]{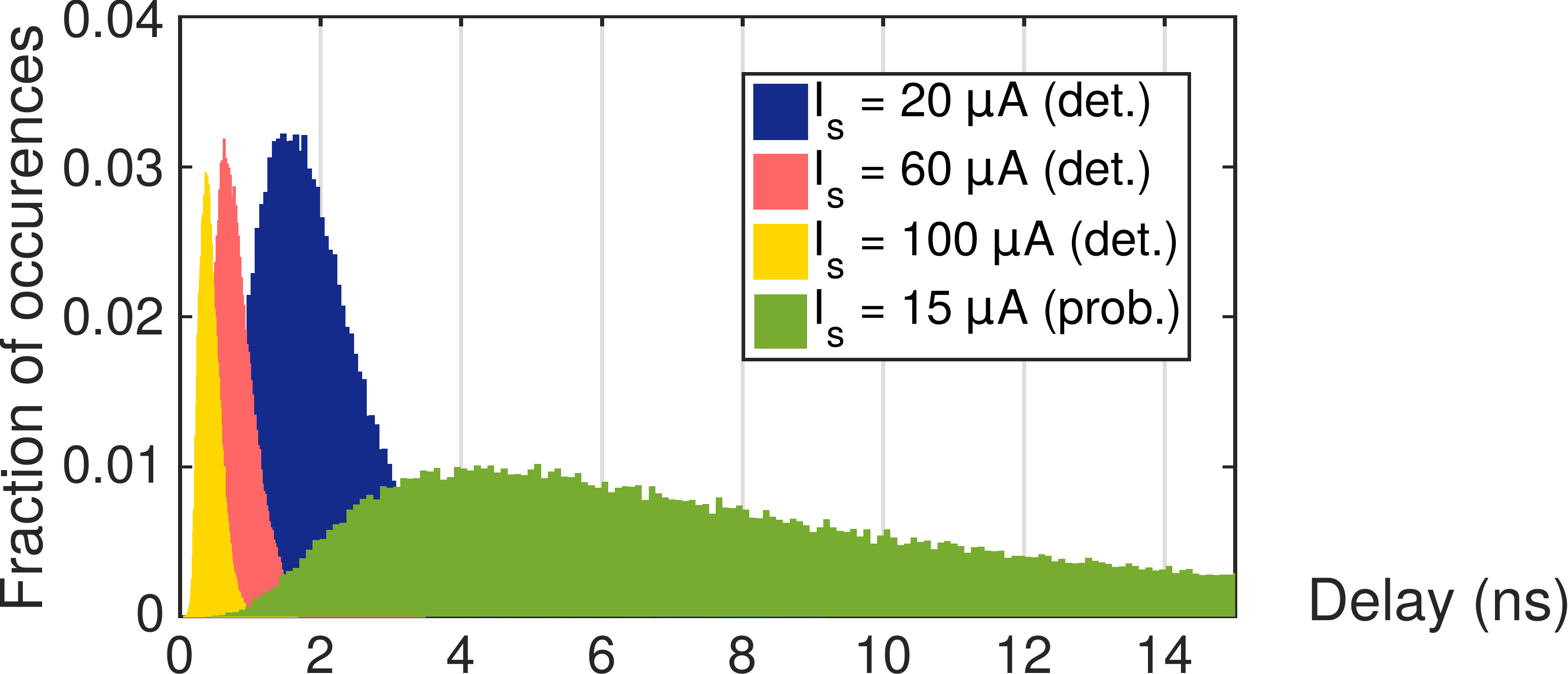}
\caption{Delay distributions for the GSHE switch at various spin currents ($I_{S}$), obtained
from 100,000 simulations each.
Note that the spread and mean delay diminish with increasing $I_{S}$, however, at the cost of
higher power dissipation.
Also note that for currents below $20$~$\mu$A, probabilistic switching occurs.
\label{fig:delay_profile}
}
  \smallerspacecaption
\end{figure}

\begin{figure}[tb]
\centering
\includegraphics[width=\textwidth]{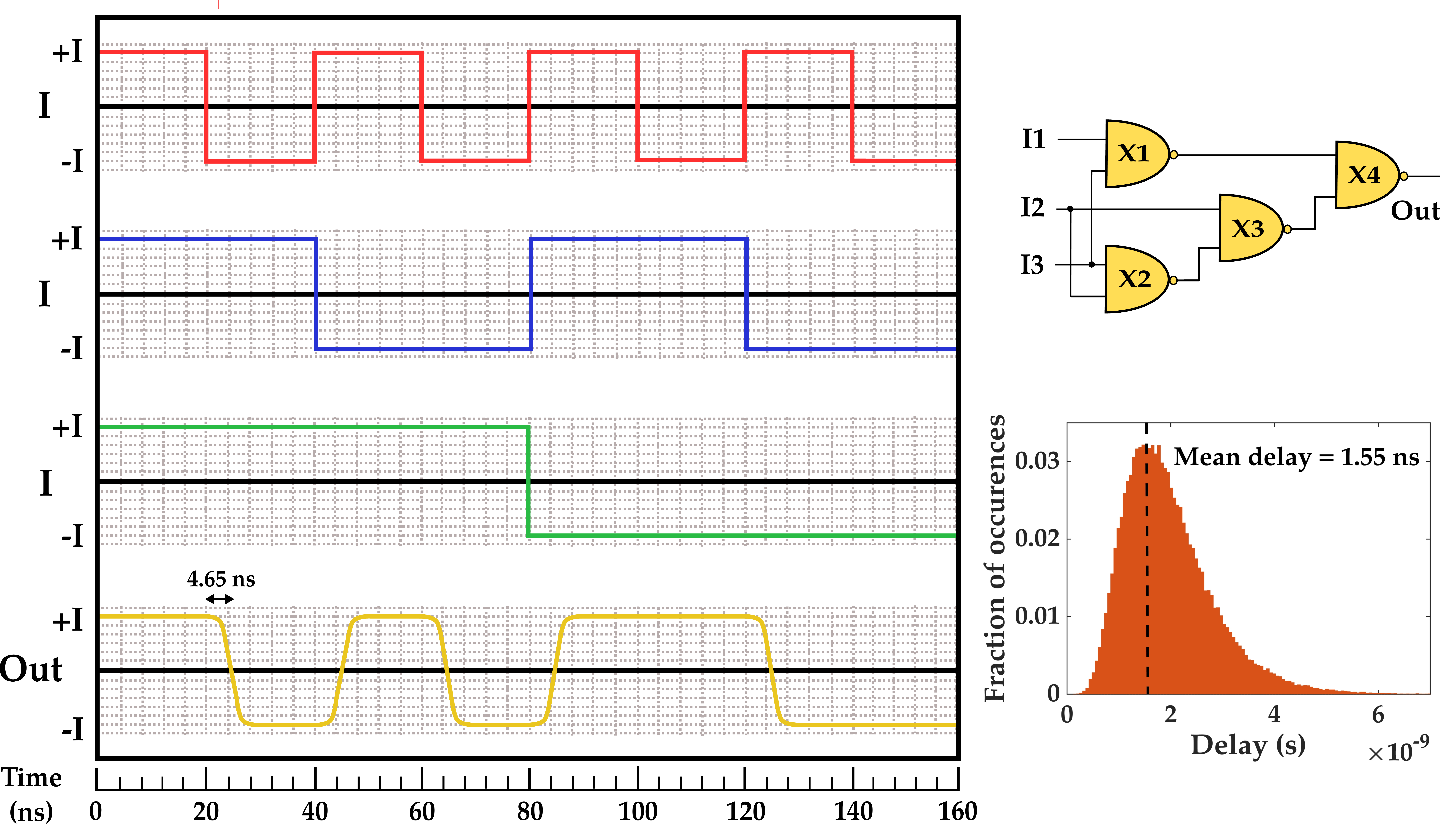}
\caption{Transient response for all input patterns applied for an example circuit (right top).
	The critical path comprises three GSHE gates, which each exhibit a mean delay of 1.55 ns (right bottom), hence the
	overall delay is 4.65 ns.
\label{fig:GSHE_timing}
}
  \smallerspacecaption
\end{figure}

The performance of the switch is determined by the nanomagnetic dynamics, which is simulated using the stochastic
Landau-Lifshitz-Gilbert-Slonczewski equation~\cite{d2006midpoint}. Simulated delay distributions
are illustrated in Fig.~\ref{fig:delay_profile}, and these distributions are computed using a CUDA-C
model~\cite{rangarajan2017energy}.
For the propagation delay for deterministic computing,
   we subsequently assume a mean delay of
1.55~ns as obtained for $I_{S} = 20$~$\mu$A, whereas for probabilistic computing, we consider a mean delay of 4.5~ns as obtained for $I_{S} = 15$~$\mu$A.
Further, this delay was then used to construct a behavioral Verilog model to obtain transient responses (Fig.~\ref{fig:GSHE_timing}).

The power dissipation
for the read-out phase is
derived according to the equivalent circuit shown in Fig.~\ref{fig:Layout_GSHE} (inset).
Using the following equations and the parameters listed in Table~\ref{tab:GSHE_parameters}, the power dissipation of the GSHE switch for
deterministic computing, including leakage, is derived as 0.2125 $\mu$W.
For probabilistic computing, power dissipation is even lower, namely $\sim$ 0.12 $\mu$W.

\begin{equation}
P = \frac{{V_{\text{OUT}}^{2}}}{r} + (V_{\text{SUP}}-V_{\text{OUT}})^{2}G_{\text{P}} + (V_{\text{OUT}}+V_{\text{SUP}})^{2}G_{\text{AP}}
\end{equation}
\begin{equation}
V_{\text{SUP}} = \left|V^{+/-}\right| = \left(\frac{I_{\text{S}}}{\beta}\right)\left(\frac{1+r(G_{\text{\text{P}}}+G_{\text{AP}})}{G_{\text{P}}-G_{\text{AP}}}\right)
\end{equation}
\begin{equation}
	V_{\text{OUT}} = \frac{I_{\text{S}}\>\> r}{\beta}
\end{equation}
\begin{equation}
\frac{G_{\text{P}}}{G_{\text{AP}}} = 1 + \text{TMR}; \; G_{\text{P}} = \frac{A(\text{nanomagnets})}{\text{RAP}}
\end{equation}

In Table~\ref{tab:devices}, we compare the GSHE switch against those of existing deterministically driven devices, including
ones that are not necessarily security-oriented.
Except for the silicon nanowire device~\cite{bi16_JETC},
the switch is superior in terms of energy/power.
When compared to \cite{bi16_JETC} (and CMOS devices, see also Sec.~\ref{sec:overheads}),
the switch has a higher delay.
In comparison to the other emerging devices, however, the delay of the GSHE switch can be considered competitive.  That is especially
justified as most prior work does not discuss their peripherals; see Sec.~\ref{sec:overheads} for our peripherals. In any case, the switch
can serve to strongly protect industrial designs without inducing significant delay overheads, as we show in Sec.~\ref{sec:GSHE-CMOS}.

As for security in terms of obfuscation, the number of possible functions is the
relevant metric---here the GSHE switch significantly outperforms most prior art.
See also Sec.~\ref{sec:overheads} for a discussion on the contending SOT-based work of~\cite{yang18}.

\begin{table}[tb]
\centering
\scriptsize
\caption{Comparison of Selected Emerging-Device Primitives
	(Deterministic Regime)
}
\label{tab:devices}
\setlength{\tabcolsep}{0.8mm}
\smallerspacecaption
\begin{tabular}{c|c|c|c|c}
\hline
\textbf{Publication} & \textbf{Functions (2 Inputs)} & \textbf{Energy} & \textbf{Power} & \textbf{Delay}\\
\hline 
\hline
~\cite{bi16_JETC} SiNW & NAND/NOR & 0.05--0.1 fJ & 1.13--1.77 $\mu$W & 42--56 ps \\ \hline
~\cite[a]{alasad2017leveraging} ASL & NAND/NOR/AND/OR & 0.58 pJ & 351.52 $\mu$W & 1.65 ns \\ \hline
~\cite[b]{alasad2017leveraging} ASL & XOR/XNOR & 1.16 pJ & 351.52 $\mu$W  & 3.3 ns \\ \hline
~\cite[c]{alasad2017leveraging} ASL & INV/BUF & 0.13 pJ & 342.11 $\mu$W & 0.38 ns \\ \hline
~\cite{huang2016magnetic} DWM  & AND/OR & 67.72 fJ & 60.46 $\mu$W & 1.12 ns \\ \hline
\multirow{2}{*}{~\cite{parveen17} DWM}  &  NAND/NOR/XOR/ & \multirow{2}{*}{N/A} & \multirow{2}{*}{N/A} & \multirow{2}{*}{N/A}	\\
&  XNOR/AND/OR/INV
	& & & \\ \hline
~\cite{zhang2015giant} GSHE  & 	AND/OR/NAND/NOR & N/A	& N/A		& N/A	\\ \hline
\multirow{2}{*}{~\cite{winograd2016hybrid} STT} &  NAND/NOR/XOR/ & \multirow{2}{*}{N/A}	& \multirow{2}{*}{N/A}		& \multirow{2}{*}{N/A}	\\
&  XNOR/AND/OR & & & \\ \hline
~\cite{yang18} SOT &  All 16 & N/A	& N/A		& N/A	\\ \hline
\hline 
\textbf{GSHE (intrinsic)} & \textbf{All 16} & \textbf{0.33 fJ} & \textbf{0.2125 $\mu$W} & \textbf{1.55 ns} \\ \hline
\textbf{GSHE + transducer} & \textbf{All 16} & \textbf{0.45 fJ} & \textbf{0.2525 $\mu$W} & \textbf{1.8 ns} \\ \hline
\textbf{Obfuscated GSHE} & \textbf{All 16} & \textbf{0.49 fJ} & \textbf{0.2673 $\mu$W} & \textbf{1.83 ns} \\ 
\textbf{(with MUXes)} &  &  & &  \\ \hline

\end{tabular}

\\[1mm]
The devices selected from prior art operate exclusively in the deterministic regime, whereas
	the GSHE switch can also operate in the probabilistic regime.
	Besides, the metrics quoted from prior art do not account for
		their peripheral circuitry.
For ours, the peripheral ME transducer is
	modeled according to~\cite{manipatruni18,iraei17}.
	Peripheral MUXes are characterized for the 15nm CMOS node using the \emph{NCSU FreePDK15 FinFET} library~\cite{bhanushali2015freepdk15},
and simulated for static camouflaging and a supply voltage of 0.75V using \emph{Cadence Virtuoso}.
	The area footprints are 0.003 $\mu$m$^2$ for the GSHE device with transducer and 0.029 $\mu$m$^2$ for the obfuscated GSHE primitive
	with all MUXes.

\end{table}

\subsection{GSHE Security Primitive: Protecting the Design IP}
\label{sec:primitive}

\begin{figure*}[tb]
    \centering
    \includegraphics[width=.9\textwidth]{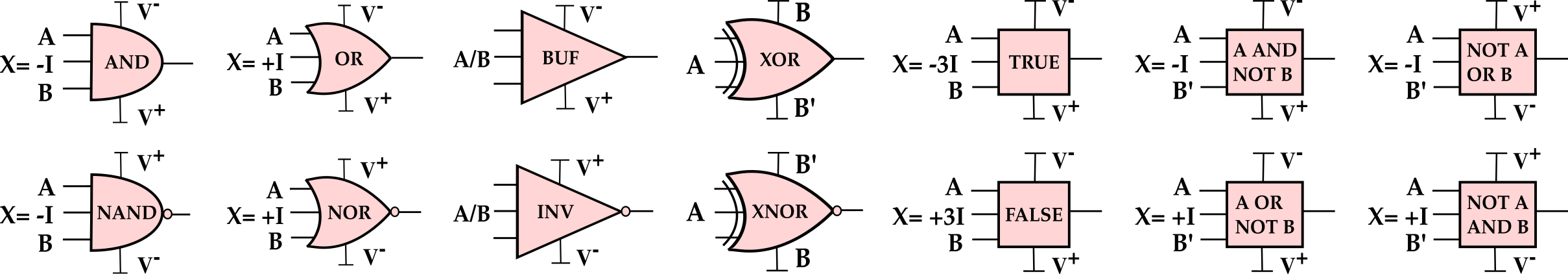}
    \caption{All 16 possible Boolean functionalities for two inputs, A and B, implemented using the proposed primitive.
	    If required, X serves as control signal, not as regular input. Note that BUF and INV capture two functionalities each.
    \label{fig:GSHE_gates}
    }
\smallerspacecaption
\end{figure*}

The GSHE switch is leveraged for a simple but versatile and effective security primitive---all 16
possible Boolean
functions can be cloaked within a single device (Fig.~\ref{fig:GSHE_gates}).
In other words, employing this primitive instead of regular gates can hinder RE attacks
of the chip's design IP, without the need for the designer to alter the underlying netlist.
For example, to realize NAND/NOR using this primitive, three charge currents are fed into the bottom layer of the GSHE switch at once
(Fig.~\ref{fig:GSHE_NAND_NOR} and \ref{fig:GSHE_gates}): two currents represent the logic signals A and B, and the third current (X) acts as the ``tie-breaking''
control input.

For some functions, the logic signals have to be provided as MTJ voltages, not as charge currents.
To transduce voltage into charge currents (as well as obtain altering current polarities), magnetoelectric
(ME) transducers can be used~\cite{manipatruni18,iraei17}. Such transducers can be placed in the interconnects, and they are capable of charge current
conversion (i.e., +I to -I) and voltage to charge current conversion (i.e., high/low voltages to +/-I) and vice versa, with relatively
	little overhead (Table~\ref{tab:devices}).

\begin{figure}[tb]
\centering
  \includegraphics[width=.95\textwidth]{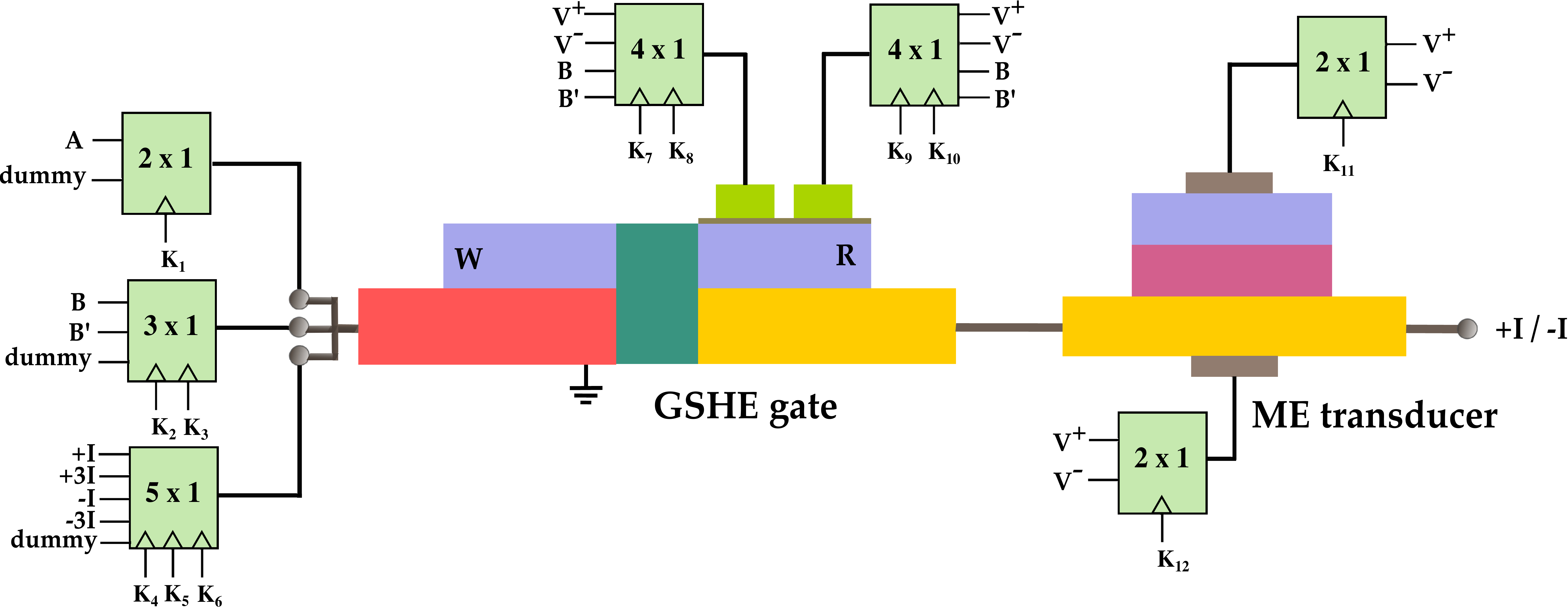}
  \caption{Peripheral circuitry (the GSHE gate is laid out horizontally for clarity).
		  For protection against fab adversaries, the MUX control signals have to connect to a tamper-proof memory or leverage
		  split manufacturing.
  \label{fig:GSHE_peripherals}
  }
\smallerspacecaption
\end{figure}

Note that three wires must be used for the GSHE input terminals (Fig.~\ref{fig:Layout_GSHE}).  This is to render the primitive indistinguishable
for imaging-based RE by malicious end-users, irrespective of the actual functionality.
Several functionalities leave some of those wires unassigned (Fig.~\ref{fig:GSHE_gates}) which 
can, e.g., be implemented as non-conductive dummy interconnects~\cite{patnaik17_Camo_BEOL_ICCAD,chen18_interconnects}, especially if only
static camouflaging is considered.
However, to support flexible input assignments, which is also essential for polymorphic switching at runtime,
we implement MUXes as peripheral circuitry for all three input wires.
(Fig.~\ref{fig:GSHE_peripherals}).
Similarly, MUXes are required for the voltage assignment for the GSHE MTJ and the ME transducer.

To hinder fab-based adversaries, we outline two equally promising options for secure implementation:
(a) leverage split manufacturing~\cite{mccants11}, (b) provision
for a tamper-proof memory.
For option (a), the MUX control signals shall remain protected from the untrusted FEOL fab.
Hence, the related wires have to be routed through the BEOL, which is then manufactured by a separate, trusted
fab~\cite{mccants11}.
Recent studies have demonstrated the practical application of split manufacturing~\cite{vaidyanathan14_2,hill13} and guided routing
within the BEOL for advancing split manufacturing~\cite{patnaik18_SM_ASPDAC,imeson13}.
For option (b), a tamper-proof memory holds a secret key that defines the correct assignment of control inputs.
The key is loaded into the memory post-fabrication by the IP holder or authorized parties.

For option (a), fab-based adversaries may employ so-called proximity attacks to try to recover the withheld BEOL routing 
from various physical-design hints present in the FEOL~\cite{rajendran13_split, wang16_sm}.
We leverage the state-of-the-art attack provided by Wang \emph{et al.}~\cite{wang16_sm}
to conduct exploratory proximity attacks on our scheme (Table~\ref{tab:CCR_results}).
Here, we note the following.
First, the correct connection rate (CCR, which quantifies the BEOL recovery) tends to
decrease for larger obfuscation scales.
For practically relevant large scales (i.e., those which cannot be resolved later on by malicious end-users employing SAT attacks, see
		Sec.~\ref{sec:security}), the attack by Wang \emph{et al.}~\cite{wang16_sm} furthermore incurs excessive runtime (R/T).
For example, for 30\% camouflaging of the benchmark \emph{b21}, when split at M6,
the attack cannot resolve within 48 hours, resulting in time-out (t-o).
Second, the CCR tends to increase for higher split layers.
Third, the larger the design, the more challenging the attack.
Overall, while some BEOL wires can be correctly inferred, proximity attacks are limited, especially when split at lower layers.
These findings are also confirmed by prior art~\cite{wang16_sm,patnaik18_SM_DAC,magana16,sengupta17_SM_ICCAD,patnaik18_SM_ASPDAC}.
Finally, also note that only we as designers, having full access to the
layout, can evaluate the attack, whereas any
fab-based adversary
can only run the attack for a ``best guess.''

\begin{table}[tb]
\centering
\scriptsize
\setlength{\tabcolsep}{0.7mm}
\caption{Exploratory Proximity Attack Results for \cite{wang16_sm}}
\smallerspacecaption
\begin{tabular}{c|c|c|c}
\hline
\textbf{Benchmark} 
& \textbf{Camouflaging (\%)} 
& \textbf{M6: CCR(\%) / R/T(s)}
& \textbf{M8: CCR(\%) / R/T(s)} \\ 
\hline \hline
ex1010 & 10 & 46 / 64 & 82 / 64 \\ \hline
ex1010 & 20 & 36 / 138 & 59 / 103 \\ \hline
ex1010 & 30 & 30 / 484 & 41 / 183 \\ \hline
c7552  & 10 & 50 / 18 & 95 / 17 \\ \hline
c7552  & 20 & 40 / 121 & 71 / 42 \\ \hline
c7552  & 30 & 29 / 136 & 47 / 72 \\ \hline
b14    & 10 & 22 / 1,457 & 41 / 141 \\ \hline
b14    & 20 & 24 / 4,513 & 33 / 387 \\ \hline
b14    & 30 & 26 / 13,293 & 37 / 777 \\ \hline
b21    & 10 & 19 / 11,504 & 38 / 567 \\ \hline
b21    & 20 & 28 / 43,646 & 40 / 1,489 \\ \hline
b21    & 30 & t-o / t-o & 28 / 8,018 \\ \hline
\hline
\textbf{Average} & -- & 32 / 6,852 & 51 / 988 \\ \hline
\end{tabular}
\label{tab:CCR_results}
\smallerspacecaption
\end{table}

Both options (a) and (b) represent a notable advancement
over prior work related to camouflaging, where
the IP holder \emph{must} trust
the fab because of the circuit-level protection mechanism.
In the remainder of this paper, we focus on the threats imposed by \emph{malicious end-users.}

\subsection{Cost and Comparison of the GSHE Security Primitive}
\label{sec:overheads}

Recall that Table~\ref{tab:devices} covers
the deterministic regime for the GSHE device and the primitive
(see also Fig.~\ref{fig:GSHE_peripherals}).
The addition of MUXes for the GSHE primitive
incurs overheads of $8.9\%$, $5.8\%$, and $1.7\%$ for energy, 
power and delay, respectively.

In Table~\ref{tab:PPA}, we further report on area (A), power (P), and delay (D) for selected benchmarks as obtained from \emph{Synopsys Design
Compiler}, using the \emph{Nangate 15nm Open Cell} CMOS library~\cite{martins2015open} (at slow corner),
the regular GSHE device (with transducer), and the GSHE security primitive, respectively.
For the deterministic GSHE implementations, the APD metrics derived above are leveraged for
full-chip gate-level mapping, i.e., each CMOS gate is assumed to be implemented as GSHE gate.
Accordingly, the results for the GSHE primitive represent the case of full-chip obfuscation.

As indicated, the GSHE implementation is inferior to regular (i.e., unprotected) CMOS implementations concerning delays.
	Considering Table~\ref{tab:devices}, this would also apply for other prior art.
Regarding area, the GSHE peripheral
MUXes induce $\sim 8\times$ additional cost compared to the regular GSHE implementation.
However, considering absolute numbers, the area for
full-chip GSHE obfuscation is, on average,
$\sim7.3\times$ smaller than for the CMOS implementation of the same benchmark. 
Here it is important to note that any camouflaging scheme 
to be applied on the unprotected CMOS implementation would further aggravate the related APD metrics.
Besides, for purely static camouflaging, the peripheral MUXes of the GSHE primitive could be omitted, and the wiring for the functional
assignment can be implemented as a mix of real and dummy interconnects.
For example, Patnaik \emph{et al.}~\cite{patnaik17_Camo_BEOL_ICCAD} have shown that such interconnect-based obfuscation can be implemented
even for full-chip camouflaging with moderate layout cost.

\begin{table}[tb]
\centering
\scriptsize
\setlength{\tabcolsep}{0.4mm}
\renewcommand{\arraystretch}{1.2}
\smallerspacecaption
\caption{Full-Chip Results for Different Implementations}
\label{tab:PPA}
\begin{tabular}{*{9}{c|}c}
\hline
& \multicolumn{3}{|c|}{}
& \multicolumn{3}{|c|}{\textbf{Regular}}
& \multicolumn{3}{|c}{\textbf{Full-Chip Obfuscated}} \\
\textbf{Benchmark}
& \multicolumn{3}{|c|}{\textbf{Regular CMOS (15nm)}} 
& \multicolumn{3}{|c|}{\textbf{Deterministic GSHE}} 
& \multicolumn{3}{|c}{
	\textbf{Deterministic GSHE}}
  \\
\cline{2-10}
 & \textbf{A($\mu$m$^{2}$)} & \textbf{P(mW)} & \textbf{D(ns)} 
 & \textbf{A($\mu$m$^{2}$)} & \textbf{P(mW)} & \textbf{D(ns)} 
 & \textbf{A($\mu$m$^{2}$)} & \textbf{P(mW)} & \textbf{D(ns)} \\ \hline
\hline

c7552 
& 222.51 & 0.68 & 0.28 
& 2.66 & 0.22 & 57.61 
& 
 26.05  
& 
 0.24  
& 
 58.74  \\ \hline

b14 
& 601.03 & 1.73 & 0.52 
& 8.49 & 0.72 & 75.6 
& 
 83.26  
& 
 0.76  
& 
 77.09  \\ \hline

b20 
& 1,283.16 & 3.56 & 0.64
& 19.07 & 1.61 & 55.8 
& 
 186.89
& 
 1.69  
& 
 56.91  \\ \hline

b21
& 1,285.62 & 2.86 & 1.02 
& 18.77 & 1.58 & 50.4 
& 
 183.92
& 
 1.67  
& 
 51.39  \\ \hline

b22
& 1,910.88 & 3.65 & 1.43 
& 28.99 & 2.44 & 72.0 
& 
 284.18  
& 
 2.58  
& 
 73.42  \\ \hline

\end{tabular}
\smallerspacecaption
\end{table}

For the recent SOT-LUT-based camouflaging scheme proposed by Yang \emph{et al.}~\cite{yang18}, we note the following.
First, to support all 16 possible functions while implementing an LUT,
their primitive requires multiple magnetic devices and related peripheral circuitry. In contrast, our switch can
implement all 16 functions within one device, with
peripheral circuitry only required for transduction and obfuscation purposes.
Second, the primitive in~\cite{yang18} is not inherently polymorphic; hence, it can only support static camouflaging.
Third, recall that the authors obfuscated only 16/32/64 gates. We found empirically that such small-scale camouflaging can
be resolved by SAT attacks, which is also confirmed by prior art~\cite{patnaik17_Camo_BEOL_ICCAD,yu17}.
Fourth, Yang \emph{et al.} report only relative area and delay overheads (power is not reported at all),
but no absolute device-level numbers (hence also the ``N/A'' labels in Table~\ref{tab:devices}).
Their relative area and delay overheads
are in the ranges of 14.23--25.8\% 
and 23.67--40.69\%, respectively, for
obfuscating \emph{only} 16--64 gates.
Considering that the authors leverage the relatively old \emph{TSMC 0.35$\mu$m} library,
the related overheads per SOT-LUT device could become prohibitive for large-scale camouflaging,
especially in the context of modern CMOS nodes (as we consider for ours).
Hence, we believe that our scheme can be superior to that of~\cite{yang18}.

\section{Security Analysis}
\label{sec:security}

Here, we elaborate on the security of the GSHE primitive
against various end-user attacks.
Most
notably, a comprehensive study for analytical SAT attacks is conducted, where we benchmark our primitive against prior art.  A key
assumption for the SAT attacks is that the GSHE primitive is applied in the context of deterministic computing---we cover the security
analysis for probabilistic computing separately in Sec.~\ref{sec:security_prob}.

\subsection{Threat Model}

The malicious end-user is interested in resolving the underlying IP implemented by the obfuscated chip. We consider that an attacker
possesses the know-how and has access to RE equipment such as required for imaging-based RE. However, the attacker does not have access to advanced
capabilities for invasive read-out attacks to, e.g., resolve the voltage and current assignments for individual GSHE primitives at runtime.
(Even if so, such attacks seem practically challenging.) 

In accordance with prior works, we assume that an attacker procures multiple copies of the chip from open market;
she/he uses one for RE (which includes de-packaging, de-layering, imaging of individual layers, stitching of these images
and final netlist extraction~\cite{quadir16}), and another as an oracle to obtain input-output (I/O)
patterns. These patterns are then utilized for SAT-based attacks.
The attacker can also use the oracle chip to evaluate side-channel leakage at runtime.

\subsection{On Reverse Engineering and Side-Channel Attacks}

\subsubsection{\textbf{Layout Identification and Read-Out Attacks}}
Recall that the physical
layout of the proposed primitive is uniform (Sec.~\ref{sec:GSHE}); hence, it remains indistinguishable for optical-imaging-based RE.
It was also shown that dummy interconnects can become difficult to resolve during RE, as long as suitable materials such as
Mg and MgO are used~\cite{hwang12_transient_electronics,chen18_interconnects}.
A more sophisticated 
attacker might, however, leverage electron microscopy (EM) for identification and read-out attacks.
For example, Courbon~\emph{et al.}~\cite{courbon16} used scanning EM, in passive voltage-contrast (PVC) mode,
to read out memories in supposedly secured chips.
While such
attacks are yet to be demonstrated on switching devices at runtime,
we believe that the proposed primitive can thwart them for three reasons.
First, the dimensions of the GSHE switch are significantly smaller than CMOS devices, which
is a challenge regarding the spatial resolution for EM-based analysis~\cite{courbon16}.
Second, the primitive readily supports probabilistic switching.  This implies that once an attacker can read-out the switch at runtime,
she/he still has to learn and account for the underlying error distributions.
Third, the primitive is truly polymorphic, and its functionality may be switched at runtime; see also next.

\subsubsection{\textbf{Polymorphism at the System Level}}
Given the truly polymorphic nature of the GSHE switch and assuming some additional circuitry to switch their functionalities judiciously,
one can implement \emph{runtime polymorphism} at the system level.
The gates' functionalities are not static anymore, possibly even for static input patterns, whereupon an attacker is bound to
misinterpret some parts of the layout---it seems impossible to resolve all dynamic features on a full-chip scale at
once.\footnote{For
example, Courbon \emph{et al.}~\cite{courbon16} report that it took 50 ns to read-out one pixel of one memory cell; this is well above
the 1.55 ns switching speed for the GSHE device for deterministic computing (recall Sec.~\ref{sec:GSHE}).}

Besides hindering read-out threats, polymorphism at the system level is also powerful to thwart SAT attacks.
In fact,
   we provide such a concept
   based on the GSHE switch 
   in Sec.~\ref{sec:PSAT_counter}.

\subsubsection{\textbf{Photonic Side-Channel Attacks}} 
It is well known that CMOS devices emit photons during operation, which makes them vulnerable to powerful attacks~\cite{tajik17_CCS,schloesser12}.
For example, Tajik \emph{et al.}~\cite{tajik17_CCS} successfully conduct an optical read-out attack against the bit-stream encryption
feature of a \emph{Xilinx Kintex 7} FPGA.
Contrary to CMOS,
the GSHE switch itself does not emit any photons.
The fundamentally different, magnetic switching principle
thus renders the primitive inherently resilient against photonic side-channel attacks.
We caution that a system-level assessment against such attacks shall be performed nevertheless (once such chips are manufactured)
	since additional circuitry may or may not remain vulnerable.

\subsubsection{\textbf{Magnetic- and Temperature-Driven Attacks}} 
Ghosh~\emph{et al.}~\cite{ghosh2016spintronics}
consider and review attacks on spintronic memory devices using external magnetic 
fields and malicious temperature curves.
As for the GSHE switch, note that it is tailored for robust magnetic coupling (between the W and R
nanomagnets)~\cite{rangarajan2017energy}, and this coupling
would naturally be disturbed by any external magnetic fields.
Hence, an attacker leveraging a magnetic probe may induce stuck-at-faults which are, however,
hardly controllable due to multiple factors: the very small size of the GSHE switch, accordingly large magnetic fields required for the probe,
the state of the nanomagnets, the orientation of the fixed magnets, and also the voltage polarities for the MTJ setup.
Temperature-driven attacks will impact the retention time of the GSHE switch.
The resulting disturbances,
however, are stochastic due to the inherent thermal noise in the nanomagnets; fault attacks are accordingly challenging as well.
As a result, we believe that subsequent sensitization attacks to resolve the obfuscated IP (e.g., as proposed
in~\cite{rajendran13_camouflage}) will be difficult, if practical at all.

\subsection{Study on Large-Scale IP Protection Against SAT Attacks}
\label{sec:det_SAT_study}

\subsubsection{\textbf{Experimental Setup}}
\label{sec:setup_det_SAT}

We model the GSHE primitive and selected prior art~\cite{rajendran13_camouflage, parveen17, alasad2017leveraging,
	zhang16, bi16_JETC, nirmala16, winograd2016hybrid, zhang2015giant} as outlined in~\cite{yu17}.
More specifically, we model the GSHE primitive as follows.
The logical inputs $a$ and $b$ are fed in parallel into all 16 possible Boolean gates, and the outputs of those gates are connecting to a
16-to-1 MUX with four select/key bits.
As for other prior art with less possible functionalities, a smaller MUX with less key bits may suffice (e.g.,
		for~\cite{rajendran13_camouflage}, a 3-to-1 MUX with two key bits is used).
Although the GSHE primitive inherently supports locking as well, here we contrast it only to camouflaging primitives, without loss of
generality.
Besides emerging-device primitives, we also contrast to CMOS-centric primitives; this is meaningful since for any IP protection scheme
the resilience against SAT attacks hinges on the number and composition of obfuscated
functionalities~\cite{massad15,subramanyan15,patnaik17_Camo_BEOL_ICCAD}, not their physical implementation.

For a fair evaluation, the same set of gates are protected; gates
are randomly selected once for each benchmark, memorized, and then the same selection is reapplied across all techniques. 
We evaluate all techniques against powerful state-of-the-art SAT attacks~\cite{subramanyan15, code_pramod, shen17},
run on
an Intel Xeon server (2.3 GHz, 4 GB per task allowed).
The time-out (labelled as ``t-o'') is set to 48 hours.
We conduct our experiments on traditional benchmarks suites, that is \emph{ISCAS-85}, \emph{MCNC}, and \emph{ITC-99}, but also on the
large-scale \emph{EPFL suite}~\cite{EPFL15} (and on the industrial \emph{IBM superblue suite}~\cite{viswanathan11}; see
		Sec.~\ref{sec:discussion} for those experiments).
The benchmarks are summarized in
Table \ref{tab:benchmarks}.

\begin{table}[tb]
\centering
\scriptsize
\caption{Characteristics of Synthesized
	Benchmarks
		(Italics: \emph{EPFL Suite}~\cite{EPFL15}; Bold: \emph{IBM Superblue Suite}~\cite{viswanathan11})
}
\label{tab:benchmarks}
\setlength{\tabcolsep}{0.7mm}
\smallerspacecaption
\begin{tabular}{c|c|c|c||c|c|c|c}
\hline
 \textbf{Benchmark
 } & \textbf{Inputs} & \textbf{Outputs} & \textbf{Gates
 } &
 \textbf{Benchmark
 } & \textbf{Inputs} & \textbf{Outputs} & \textbf{Gates 
 } \\ \hline \hline
\emph{aes\_{core}}  & 789 & 668 & 39,014 & \emph{log2} & 32 & 32 & 51,627 \\ \hline
b14 & 277 & 299 & 11,028 & \textbf{sb1} & 8,320 & 13,025 & 856,403 \\ \hline
b21 & 522 & 512 & 22,715 & \textbf{sb5} & 11,661 & 9,617 & 741,483 \\ \hline
c7552 & 207 & 108 & 4,045 & \textbf{sb10} & 10,454 & 23,663 & 1,117,846 \\ \hline
ex1010 & 10 & 10 & 5,066 & \textbf{sb12} & 1,936 & 4,629 & 1,523,108 \\ \hline
\emph{pci\_bridge32} & 3,520 & 3,528 & 35,992 & \textbf{sb18} & 3,921 & 7,465 & 659,511 \\ 
\hline
\end{tabular}

\smallerspacecaption
\end{table}

\subsubsection{\textbf{Results}}
In Table~\ref{tab:SAT_pramod}, we report the runtimes incurred by the seminal attack of Subramanyan \emph{et al.}~\cite{subramanyan15,
code_pramod}. While there are further metrics such as the number of clauses, attack iterations, or number of remaining feasible
assignments~\cite{yu17}, runtime is a straightforward yet essential indicator---either an attack succeeds within the allocated time 
		or not.

We observe that for the same number of gates protected, the more functions a primitive can cloak, the more resilient it becomes in practice.
More importantly, the runtimes required for decamouflaging---if possible at all---tend to scale exponentially with the percentage of gates
being camouflaged.\footnote{Inducing prohibitive computational cost is also the primary objective for provably secure schemes as
in~\cite{yasin16_SARLock, xie16_SAT, li16_camouflaging}. We further elaborate on provably secure schemes versus our large-scale scheme in
Sec.~\ref{sec:provably_vs_large}.}

\begin{table*}[tb]
\centering
\scriptsize
\setlength{\tabcolsep}{1mm}
\caption{Runtime for SAT Attacks \cite{subramanyan15,code_pramod},
	on Selected Designs,
	in Seconds (Time-Out ``t-o'' is 48 Hours, i.e., 172,800 Seconds)
}\label{tab:SAT_pramod}
\smallerspacecaption
\begin{tabular}{*{14}{c|}c}
\hline
\multirow{3}{*}{\textbf{Benchmark}}
& \multicolumn{7}{|c|}{\textbf{10\% IP Protection}} & \multicolumn{7}{|c}{\textbf{20\% IP Protection}} \\
\cline{2-15}
& \textbf{\cite{rajendran13_camouflage}} & \textbf{\cite{nirmala16,winograd2016hybrid}}
& \textbf{\cite{bi16_JETC}}
	& \textbf{\cite[c]{alasad2017leveraging}, \cite{zhang16}} & \textbf{\cite{zhang2015giant}, \cite[a]{alasad2017leveraging}} &
	\textbf{\cite{parveen17}} & \textbf{Our}
& \textbf{\cite{rajendran13_camouflage}} & \textbf{\cite{nirmala16,winograd2016hybrid}}
& \textbf{\cite{bi16_JETC}}
	& \textbf{\cite[c]{alasad2017leveraging}, \cite{zhang16}} & \textbf{\cite{zhang2015giant}, \cite[a]{alasad2017leveraging}} &
	\textbf{\cite{parveen17}} & \textbf{Our}
\\
& (3)$^*$
& (6)$^*$
& (4)${^*}{^\dag}$
& (2)$^*$
& (4)$^*$
& (7+1)${^*}{^\ddag}$
& (16)$^*$
& (3)$^*$
& (6)$^*$
& (4)${^*}{^\dag}$
& (2)$^*$
& (4)$^*$
& (7+1)${^*}{^\ddag}$
& (16)$^*$
\\ \hline
\hline
 
\emph{aes\_{core}}
& 610 & 4,710  & 890 & 132 & 536 & 6,229 & 25,890
& 4,319 & 41,844 & 11,306 & 407 & 9,432 & t-o & t-o
\\ \hline

b14
& 2,078 & 20,603 & 11,465 & 6,884  & 17,634 & 27,438 & 60,306
& 56,155 & t-o & 64,145 & 8,426 & t-o & t-o & t-o
\\ \hline

b21
& 7,813 & 162,324 & 45,465 & 3,977  & 24,035 & t-o & t-o
& t-o & t-o & t-o & t-o  & t-o & t-o & t-o
\\ \hline

c7552
& 37 & 210 & 74 & 12  & 66 & 371 & 2,289
& 169 & 14,575 & 1,153 & 110  & 1,327 & 172,548 & t-o
\\ \hline

ex1010
& 62 & 215 & 82 & 12  & 73 & 295 & 922
& 171 & 1,047 & 274 & 38  & 250 & 1,310 & 4,701
\\ \hline

\emph{log2}
& t-o & t-o & t-o & t-o  & t-o & t-o & t-o
& t-o & t-o & t-o & t-o  & t-o & t-o & t-o
\\ \hline

\emph{pci\_bridge32}
& 1,119 & t-o & 9,011 & 1,325  & 2,690 & t-o & t-o
& 54,577 & t-o & t-o & t-o  & t-o & t-o & t-o
\\
\hline

& \multicolumn{7}{|c|}{\textbf{30\% IP Protection}} &
	\multicolumn{7}{|c}{\textbf{40--100\% IP Protection$^\S$}}  \\
\hline
\hline
 
\emph{aes\_{core}}
& 17,148 & t-o & 31,601 & 2,020 & 26,498 & t-o & t-o & 
t-o & t-o & t-o & 8,206 & t-o & t-o & t-o
\\ \hline

b14
& 56,787 & t-o & t-o & 38,495 & t-o & t-o & t-o
& t-o & t-o & t-o & t-o & t-o & t-o & t-o
\\ \hline

b21
& t-o & t-o & t-o & t-o & t-o & t-o & t-o
& t-o & t-o & t-o & t-o & t-o & t-o & t-o
\\ \hline

c7552
& 1,786 & t-o & t-o & 766 & t-o & t-o & t-o
& t-o & t-o & t-o & 41,721 & t-o & t-o & t-o
\\ \hline

ex1010
& 448 & 4,357 & 938 & 87 & 719 & 11,736 & 24,727
& 1,703 & t-o & 129,290 & 169---7,073$^\S$  & 1,950 & t-o & t-o
\\ \hline

\emph{log2}
& t-o & t-o & t-o & t-o & t-o & t-o & t-o
& t-o & t-o & t-o & t-o & t-o & t-o & t-o
\\ \hline

\emph{pci\_bridge32}
& t-o & t-o & t-o & t-o & t-o & t-o & t-o
& t-o & t-o & t-o & t-o & t-o & t-o & t-o
\\ \hline

\end{tabular}

\\[1mm]
$^*$Number of cloaked functions; refer to Table~\ref{tab:devices} or the related publication for the actual sets of cloaked functions.
Prior art covering the same set is grouped into one column.
$^\dag$Here we refer to their camouflaging primitive, not the polymorphic gate reported on in Table~\ref{tab:devices}.
$^\ddag$Here
we also assume BUF to be available.
$^\S$The benchmark \emph{ex1010} can be resolved even for 100\% IP protection, but only for the primitives
of~\cite[c]{alasad2017leveraging}, \cite{zhang16}. The related runtime range is for 40--100\% protection, whereas all other runtimes are for
40\% protection (50\% protection or more ran into timeout).

\end{table*}

\begin{table*}[tb]
\centering
\scriptsize
\setlength{\tabcolsep}{1mm}
\caption{Runtime for \emph{Double DIP} Attacks \cite{shen17},
	on Selected Designs,
	in Seconds (Time-Out ``t-o'' is 48 Hours, i.e., 172,800 Seconds)
}\label{tab:SAT-double-DIP}
\smallerspacecaption
\begin{tabular}{*{14}{c|}c}
\hline
\multirow{3}{*}{\textbf{Benchmark}}
& \multicolumn{7}{|c|}{\textbf{10\% IP Protection}} & \multicolumn{7}{|c}{\textbf{20\% IP Protection}} \\
\cline{2-15}
& \textbf{\cite{rajendran13_camouflage}} & \textbf{\cite{nirmala16,winograd2016hybrid}}
& \textbf{\cite{bi16_JETC}}
	& \textbf{\cite[c]{alasad2017leveraging}, \cite{zhang16}} & \textbf{\cite{zhang2015giant}, \cite[a]{alasad2017leveraging}} &
	\textbf{\cite{parveen17}} & \textbf{Our}
& \textbf{\cite{rajendran13_camouflage}} & \textbf{\cite{nirmala16,winograd2016hybrid}}
& \textbf{\cite{bi16_JETC}}
	& \textbf{\cite[c]{alasad2017leveraging}, \cite{zhang16}} & \textbf{\cite{zhang2015giant}, \cite[a]{alasad2017leveraging}} &
	\textbf{\cite{parveen17}} & \textbf{Our}
\\
& (3)$^*$
& (6)$^*$
& (4)${^*}{^\dag}$
& (2)$^*$
& (4)$^*$
& (7+1)${^*}{^\ddag}$
& (16)$^*$
& (3)$^*$
& (6)$^*$
& (4)${^*}{^\dag}$
& (2)$^*$
& (4)$^*$
& (7+1)${^*}{^\ddag}$
& (16)$^*$
\\ \hline
\hline
 
\emph{aes\_{core}}
& 1,814 & 27,274 & 3,039 & 431 & 2,103 & 22,936 & 53,434 
& 24,635 & t-o & 55,699 & 1,631 & 34,040 & t-o & t-o
\\ \hline

b14
& 4,866 & 47,303 & 8,197 & 344 & 8,299 & 52,657 & t-o 
& 62,698 & t-o & 138,809 & 4,757 & t-o & t-o & t-o
\\ \hline

b21
& 14,671 & t-o & 84,483 & 2,095 & 47,937 & t-o & t-o 
& t-o & t-o & t-o & t-o & t-o & t-o & t-o
\\ \hline

c7552
& 58 & 763 & 153 & 19 & 173 & 1,919 & 23,632 
& 639 & t-o & 31,485 & 199 & 111,580 & t-o & t-o
\\ \hline

ex1010
& 126 & 470 & 194 & 27 & 153 & 627 & 1,897 
& 396 & 3,280 & 628 & 94 & 560 & 4,361 & 11,660
\\ \hline

\emph{log2}
& t-o & t-o & t-o & t-o & t-o & t-o & t-o
& t-o & t-o & t-o & t-o & t-o & t-o & t-o
\\ \hline

\emph{pci\_bridge32}
& 5,389 & t-o & t-o & 6,888 & t-o & t-o & t-o
& t-o & t-o & t-o & t-o & t-o & t-o & t-o
\\ \hline

& \multicolumn{7}{|c|}{\textbf{30\% IP Protection}} &
	\multicolumn{7}{|c}{\textbf{40--100\% IP Protection$^\S$}}  \\
\hline
\hline
 
\emph{aes\_{core}}
& 59,487 & t-o & t-o & 14,497 & t-o & t-o & t-o & 
t-o & t-o & t-o & 28,228 & t-o & t-o & t-o
\\ \hline

b14
& t-o & t-o & t-o & 39,128 & t-o & t-o & t-o
& t-o & t-o & t-o & t-o & t-o & t-o & t-o
\\ \hline

b21
& t-o & t-o & t-o & t-o & t-o & t-o & t-o
& t-o & t-o & t-o & t-o & t-o & t-o & t-o
\\ \hline

c7552
& t-o & t-o & t-o & 22,521 & t-o & t-o & t-o
& t-o & t-o & t-o & t-o & t-o & t-o & t-o
\\ \hline

ex1010
& 1,247 & 17,143 & 3,305 & 192 & 1,842 & 60,970 & t-o
& 8,226 & t-o & 102,512 & 396---42,543$^\S$ & 7,120 & t-o & t-o
\\ \hline

\emph{log2}
& t-o & t-o & t-o & t-o & t-o & t-o & t-o
& t-o & t-o & t-o & t-o & t-o & t-o & t-o
\\ \hline

\emph{pci\_bridge32}
& t-o & t-o & t-o & t-o & t-o & t-o & t-o
& t-o & t-o & t-o & t-o & t-o & t-o & t-o
\\ \hline

\end{tabular}

\\[1mm]
		Refer to Table~\ref{tab:SAT_pramod} for footnotes.
\end{table*}

In comparison with prior art, our primitive induces by far the largest efforts across all benchmarks. Except for \emph{ex1010}, none of the
benchmarks could be resolved
within 48 hours once we protect 20\% or more of all gates.  To confirm this superior resilience of our primitive, we conducted further
experiments running for 240 hours with full-chip protection (100\% camouflaging)---the designs could still not be resolved.
Moreover, we also observe some computational failures (e.g., ``\emph{internal error in `lglib.c': more than 134,217,724 variables}'');
this hints at practical limitations about the scalability of SAT attacks, as one can reasonably expect~\cite{massad15}.

We also apply \emph{Double DIP}, provided by Shen \emph{et al.}~\cite{shen17}.
The key advancement of their attack is that it rules out at least two incorrect keys in each iteration.
Conducting the very same set of experiments as before,
we observe that the runtimes are on average even higher across all benchmarks (Table~\ref{tab:SAT-double-DIP}).
For example, decamouflaging the benchmark \emph{aes\_core} when 10\% IP protection is applied using our primitive requires $\approx$7
hours for~\cite{subramanyan15}, but $\approx$15 hours for~\cite{shen17}.
This implies that \emph{Double DIP}, while successful for protection schemes such as \emph{SARLock}~\cite{yasin16_SARLock},
cannot cope well with our large-scale camouflaging scheme.
While \emph{Double-DIP} 
is normally used to reduce a compound defense technique 
(for e.g., \emph{SARLock} + \emph{SLL}~\cite{yasin16_SARLock}) to its 
low-error-rate constituent by ``peeling off'' the high-error-rate constituent, we apply it on our technique anyways for the sake 
of a complete security analysis.

\subsubsection{\textbf{On Provably Secure Versus Large-Scale Schemes}}
\label{sec:provably_vs_large}

Contrary to provably secure schemes such as~\cite{yasin16_SARLock, xie16_SAT, li16_camouflaging},
which are often backed by mathematical formulations,
one may find it difficult to engage in ``plain'' but large-scale camouflaging.
The reason is that the solution space $C$, covering all possible functionalities of the design after camouflaging, is
hard to quantify precisely~\cite{massad15,subramanyan15,li16_camouflaging}.
More specifically, $C$ depends 
on (i)~the number and composition of functions cloaked by each primitive,
(ii)~the number of gates protected,
(iii)~the selection of gates protected,
and (iv)~the interconnectivity of the design.
Since all aspects are interacting, SAT attacks may or may not be able to prune $C$ efficiently,
but this can only be evaluated by running the attacks.
As shown above for~\cite{shen17}, some heuristics can, in fact, be counterproductive when tackling large-scale camouflaging.
Also recall that prior schemes are limited in both (i) and (ii) by cost considerations
(Sec.~\ref{sec:prior_art}).\footnote{A massive interconnectivity may also impose a substantial cost, but more importantly, most prior
	studies ignore the potential of obfuscating the interconnects for IP protection to begin with. Patnaik \emph{et
		al.}~\cite{patnaik17_Camo_BEOL_ICCAD} proposed a dedicated design flow for low-cost and large-scale obfuscation of interconnects.}
In contrast, thanks to the innate polymorphism of the proposed GSHE primitive, we are unbound toward large-scale camouflaging with
all 16 possible functionalities cloaked within one device.

As a result, we believe that our scheme can be competitive against
provably secure techniques. In this context it is important to also note that
provably secure schemes have to trade-off corruptibility
and resilience against SAT
attacks~\cite{shamsi18_TIFS}: the larger the desired resilience, the lower the corruptibility, and vice versa.
This trade-off implies that a high-resilience scheme comes at the cost of effectively protecting only a small part of the
IP.  For our scheme, however, these concerns are inherently mitigated. That is, we can readily protect all of the IP, and the
resilience of our schemes
relies on incurring an excessive computational cost for the SAT solver (as also discussed
in~\cite{massad15,shamsi18}), not on low corruptibility.

Overall, we are not claiming that large-scale camouflaging
cannot be resolved eventually using SAT or other attacks,
although limits for computation will remain in any case~\cite{massad15}.
Rather, we provide strong empirical evidence
that attacking schemes as ours incurs prohibitive computational cost.

\section{Security Analysis for the Probabilistic Regime}
\label{sec:security_prob}

So far we have leveraged the GSHE primitive in the context of classical, deterministic computation.
In this section, we explore the implications of probabilistic computing for IP protection, which have been largely ignored until
now.

\subsection{Conventional SAT Attacks on Probabilistic Circuits}
\label{sec:SAT_attack_fail}

Consider an obfuscated, hybrid circuit in which some of the gates are probabilistic GSHE gates, while the rest are either implemented in CMOS or
as deterministic GSHE gates (see also Sec.~\ref{sec:discussion}).
Any probabilistic GSHE gate might function erroneously at any point in time, possibly corrupting the overall circuit output.
When such a probabilistic circuit is leveraged as an oracle, it might
not always faithfully produce the originally intended I/O patterns.
Which of the individual patterns is erroneous, however, remains incomprehensible to an attacker who has
yet to resolve the obfuscated functionality of the circuit.
Without a deterministic oracle to prune only the incorrect keys, SAT attacks
may observe conflicting hints or falsely prune the correct
key---a conventional SAT attack inevitably tends to fail for probabilistic circuits. 

As for a simple example, consider \emph{c17} of Fig.~\ref{fig:c17_example} again, but assume that the gate $X_6$ is being
replaced by a probabilistic GSHE NAND gate with an error rate of 5\%.
Since $X_6$ impacts the
primary output $O_2$,
the oracle will produce correct outputs only for 95\% of all inputs.
Now, for a conventional SAT attack, such I/O errors can result in false pruning of keys (Table~\ref{tab:SAT_attack_example2}).

\begin{table*}[tb]
\scriptsize
\caption{SAT Attack on the Benchmark \emph{c17}, Locked as in Fig.~\ref{fig:c17_example}, but with Gate $X_6$ Now Acting
	Probabilistically, with 5\% Error Rate
\label{tab:SAT_attack_example2}
}
\centering
\smallerspacecaption
\begin{tabular}{ P{1.5cm} | P{1.56cm} | P{0.5cm} | P{0.5cm} | P{0.5cm} | P{0.5cm} | P{0.5cm} | P{0.5cm} | P{0.5cm} | P{0.5cm} | P{6.2cm} }
\hline
{\textbf{Input patterns}} & \textbf{Oracle Output 0.95\%/0.05\%}& \multicolumn{8}{c|}{{\multirow{2}{*}{\textbf{Most Probable Output for
	Different Key Combinations}}}} &{\multirow{2}{*}{\textbf{Inference}}}\\
\cline{2-11}
$\bm{\text{\textbf{I}}_1\text{\textbf{I}}_2\text{\textbf{I}}_3\text{\textbf{I}}_4\text{\textbf{I}}_5}$&$\bm{\text{\textbf{O}}_1\text{\textbf{O}}_2}$ &$\bm{k_0}$ &$\bm{k_1}$ &$\bm{k_2}$ &$\bm{k_3}$ &$\bm{k_4}$ &$\bm{k_5}$ &$\bm{k_6}$ &$\bm{k_7}$ &\\
\hline
\hline
$00000$ &$00/01$  &$01$ &$00$ &$10$ &$11$ &$00$ &$01$ &$10$ &$11$ & \\
\hline
$00001$ &$01/00$  &$00$ &$01$ &$10$ &$11$ &$01$ &$00$ &$10$ &$11$ & \\
\hline
$00010$ &$11/10$  &$10$ &$11$ &$01$ &$00$ &$10$ &$11$ &$00$ &$01$ & \\
 \hline
 $00011$ &$11/10$  & $10$&$11$ &$00$ &$01$ &$10$ &$11$ &$01$ &$00$ & \\
\hline
$00100$ &$00/\colorbox{yellow}{01}$ &$01$ &\cellcolor{red!25}$00$ &\cellcolor{red!25}$10$ &\cellcolor{red!25}$11$ &\cellcolor{red!25}$00$
&$01$ &\cellcolor{red!25}$10$ &\cellcolor{red!25}$11$ &Iteration 1: probabilistic oracle assumes incorrect output $\Rightarrow k_1,
	k_2, k_3, k_4, k_6, k_7$ are (falsely) pruned \\
\hline
$00101$ &$01/00$  &$00$ &$01$ &$10$ &$11$ &$01$ &$00$ &$10$ &$11$ & \\
\hline
$00110$ &$11/00$  &$10$ &$11$ &$01$ &$00$ &$10$ &$11$ &$00$ &$01$ & \\
\hline
$00111$ &$\colorbox{yellow}{11}/00$ &$\cellcolor{red!25}10$ &$11$ &$00$ &$01$ &$10$ &\cellcolor{green!25}$11$ &$01$ &$00$ &Iteration 2:
probabilistic oracle assumes correct output $\Rightarrow k_1$
pruned $\Rightarrow k_5$ is inferred as (only seemingly) correct key \\
\hline
$\dots$ &$\dots$ &$\dots$ &$\dots$ &$\dots$ &$\dots$ &$\dots$ &$\dots$ &$\dots$ &$\dots$ & \\
\hline
$11111$ &$10/11$  &$11$ &$10$ &$10$ &$11$ &$11$ &$10$ &$10$ &$11$ & \\
\hline
\end{tabular}

\\[1mm]
	As in Table~\ref{tab:SAT_attack_example}, the labels $k_0$--$k_7$ represent all possible combinations of key bits, from
	$000$ to $111$, and the columns denote the corresponding, most probable outputs, which are compared with the oracle output.
	The oracle output is now probabilistic.
\end{table*}

\begin{figure*}[tb]
\centering
{
\subfloat[Circuit \emph{c432}]{
\includegraphics[width=.47\textwidth]{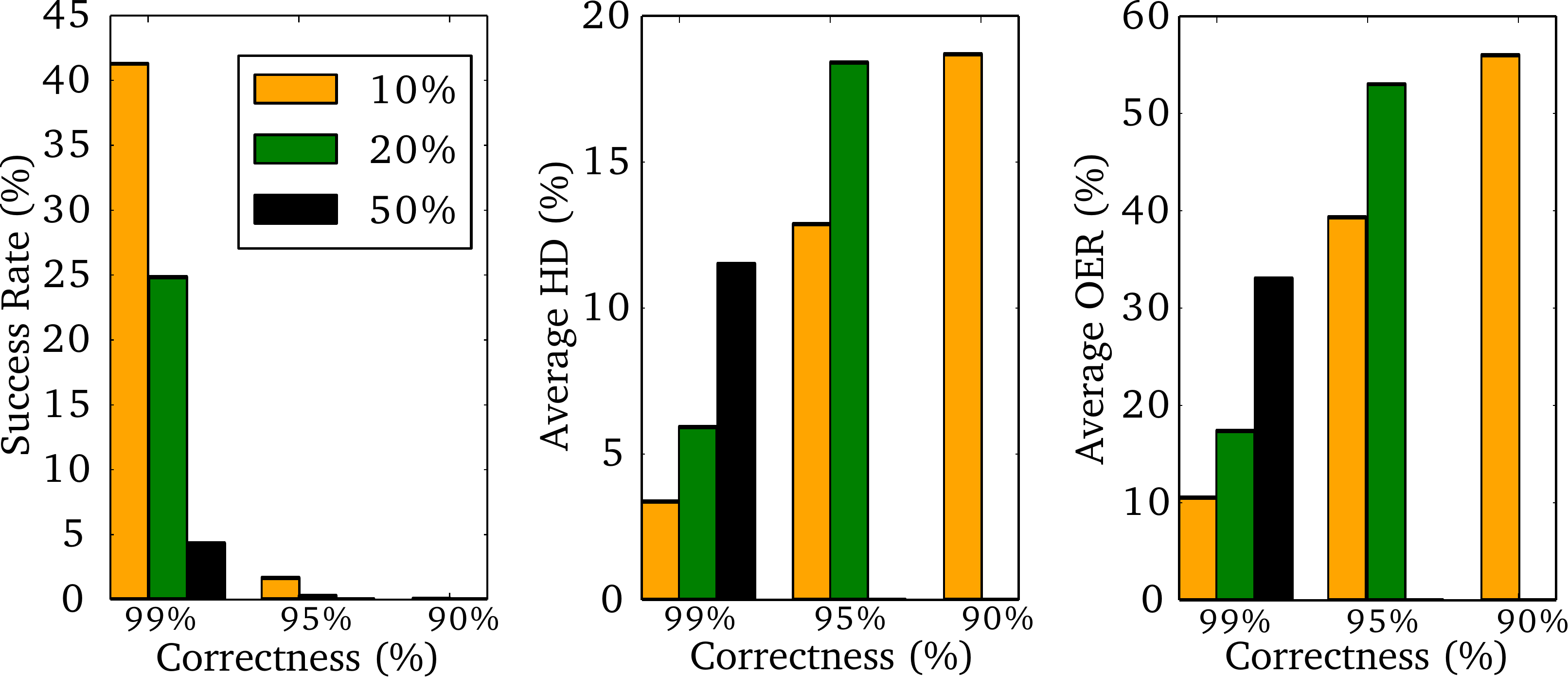}
}
\hfill
\subfloat[Circuit \emph{c880}]{
\includegraphics[width=.47\textwidth]{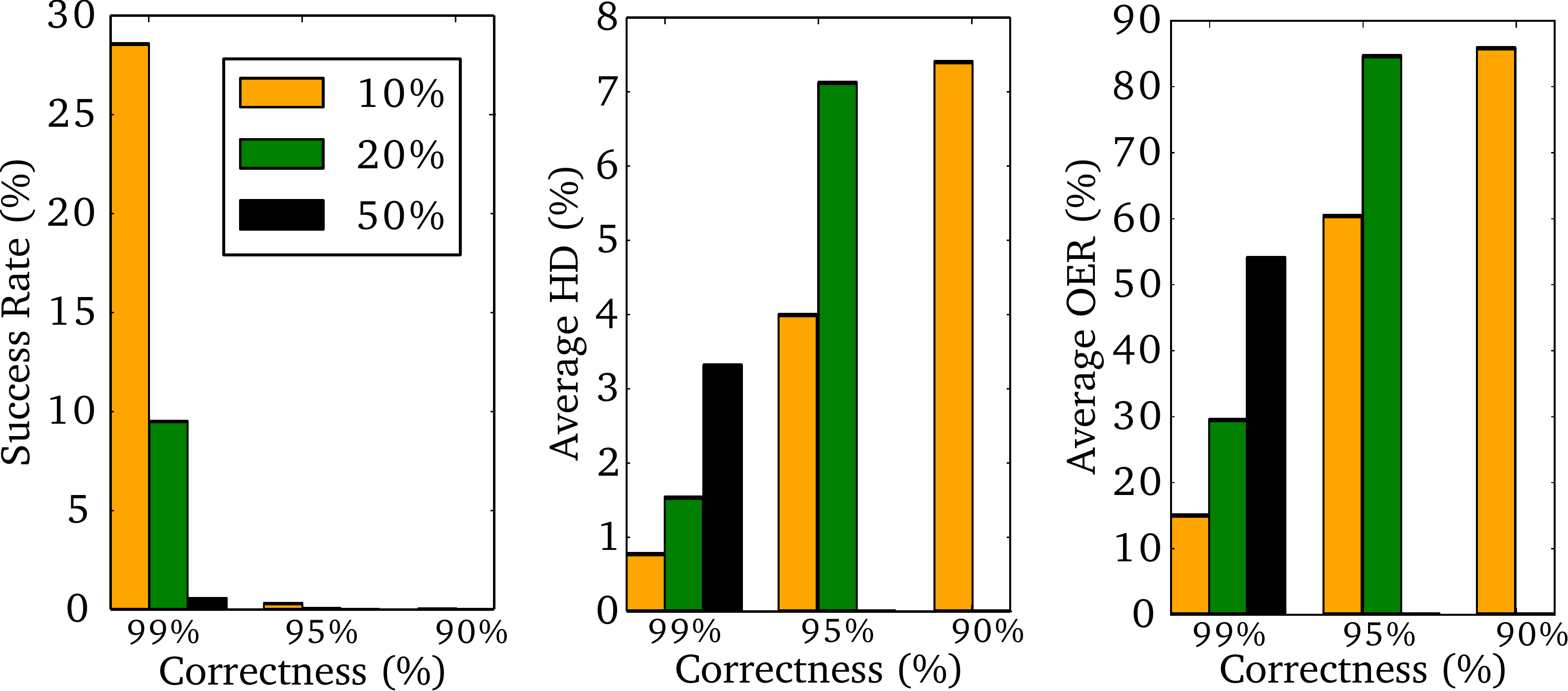}
}
\caption{Success rate, average HD, and average OER for 10,000 runs of conventional SAT
	attack~\cite{code_pramod,webinterface} applied on probabilistic versions of the \emph{ISCAS-85} benchmarks.
		The legend applies to all plots, and it represents the range of randomly selected probabilistic gates: 10\%, 20\%, and 50\% of
		all gates, respectively. The random selection is re-applied for all setups (including those in 
				Fig.~\ref{fig:PSAT_prob_circuit}), for a fair comparison.
		Correctness is the inverse of the gate error rate; for each setup, the respective correctness is constant for all
		probabilistic gates.  For cases where the attack success rate is zero, HD and OER are not available.
\label{fig:conv_attack_prob_circuit}
}
    }
\smallerspacecaption
\end{figure*}

\subsubsection{\textbf{Experimental Setup}}
\label{sec:setup_prob_SAT}

To verify this intuition, we prepare the following experiments
(with the setup described in Sec.~\ref{sec:setup_det_SAT} serving as a general baseline).
Without loss of generality, we pick \emph{c432} and \emph{c880} from the \emph{ISCAS-85} benchmark suite
and \emph{apex4} and \emph{des} from the \emph{MCNC} suite.
On those circuits, we apply a simple obfuscation scheme, namely a random insertion of 32 XOR/XNOR key gates.
Note that such simple obfuscation can be readily
resolved for deterministic circuits when using any SAT attack, especially for such relatively small circuits.
Next, we compile probabilistic versions for those obfuscated circuits.
To do so, we randomly select, memorize, and replace fixed subsets of gates (50\%, 20\%, and 10\% of all gates)
with probabilistic GSHE gates.
We model the probabilistic gate behavior directly within the conventional SAT attack by Subramanyan \emph{et al.}~\cite{subramanyan15}, whose
open-source framework~\cite{code_pramod} allows for such customization.
Note that we provide our modification of~\cite{code_pramod}, along with the probabilistic benchmark versions, as
	open source as well~\cite{webinterface}.

We run our modified but conventional SAT attack~\cite{webinterface} 10,000 times each on three different setups for the
probabilistic circuits, assuming error rates of 1\%, 5\%, and 10\%, respectively. For
simplicity, we assign identical error rates for all the probabilistic gates.\footnote{Using our setup~\cite{webinterface}, one can, however, apply different error
rates for different gates, if considered useful for the design under consideration.}
Whenever an attack is successful,
we also evaluate the Hamming distance (HD) and output error rate (OER)
between the outputs of the original but probabilistic circuit and the probabilistic circuit as
resolved after the attack. We apply 10,000 random patterns for averaging the HD and OER (here and for all subsequent setups).
It is important to note that HD and OER
provide a combined measure of error for
both the key inferred by SAT attacks and the probabilistic nature of the circuits under consideration.

\subsubsection{\textbf{Results}}

The outcome of these experiments is presented in Fig.~\ref{fig:conv_attack_prob_circuit}.
Besides the aspects discussed in Sec.~\ref{sec:provably_vs_large}, here we further note that the success rate for the conventional SAT attack
depends on (i) the number of probabilistic gates, (ii) the type/function of those gates, and (iii) their error rates.
As expected, the success rate decreases and the average HD inflates once the error rates are ramped up.  In fact, for some cases with error
rates of 5\% or 10\%, the success rate is even zero (and for such cases, HD and OER cannot be calculated).
In short, when tackling obfuscated probabilistic circuits, the capabilities of a conventional SAT attack
are drastically reduced once the correctness of the circuit is lowered.

This finding clearly illustrates the 
trade-off between correctness/accuracy and security. (Also recall Sec.~\ref{sec:case_study_app} and Sec.~\ref{sec:characterization}
		for the trade-off between accuracy and power.)
Hence, whenever the designer seeks to
forgo accuracy (typically to reduce power), the resilience against conventional SAT attacks is inherently boosted at the same time.

\begin{figure*}[tb]
\centering
{
\subfloat[Circuit \emph{c432}]{
\includegraphics[width=.47\textwidth]{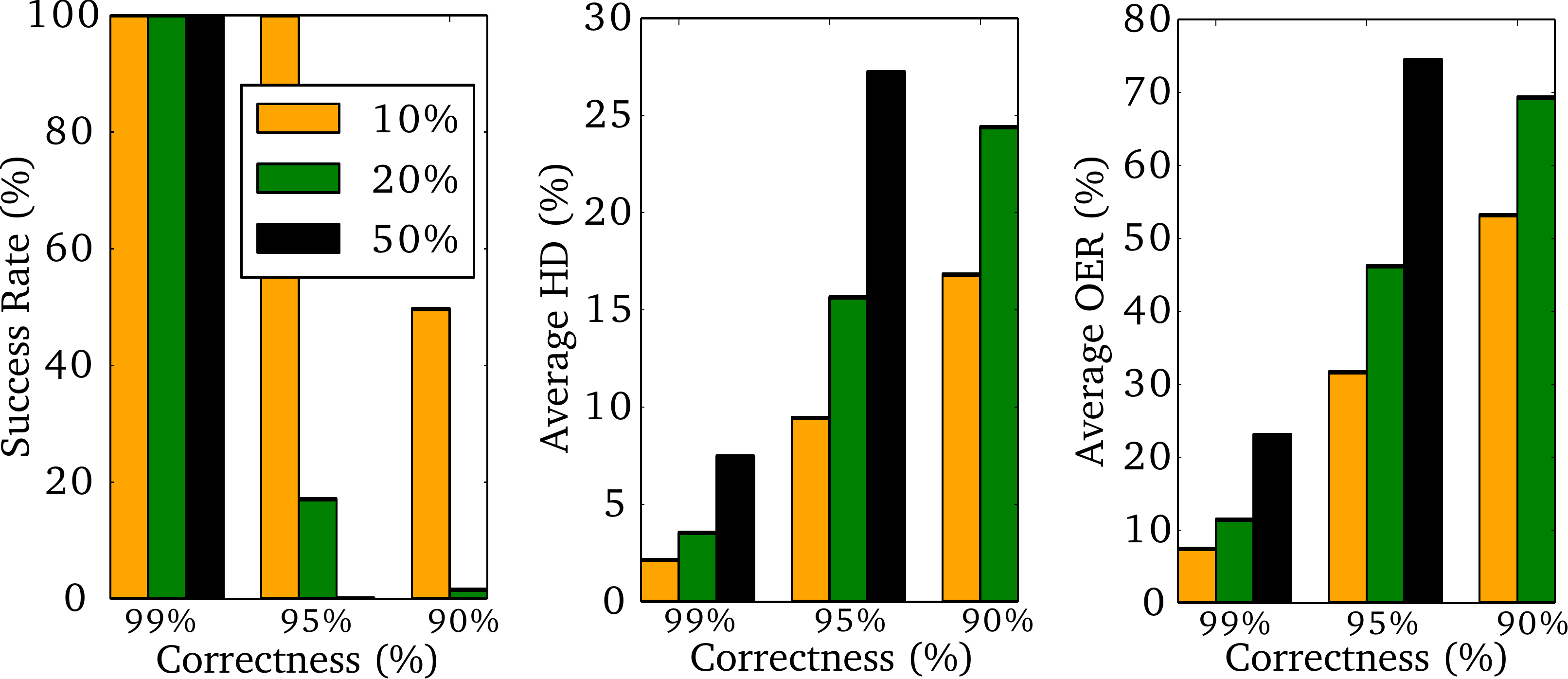}
}
\hfill
\subfloat[Circuit \emph{c880}]{
\includegraphics[width=.47\textwidth]{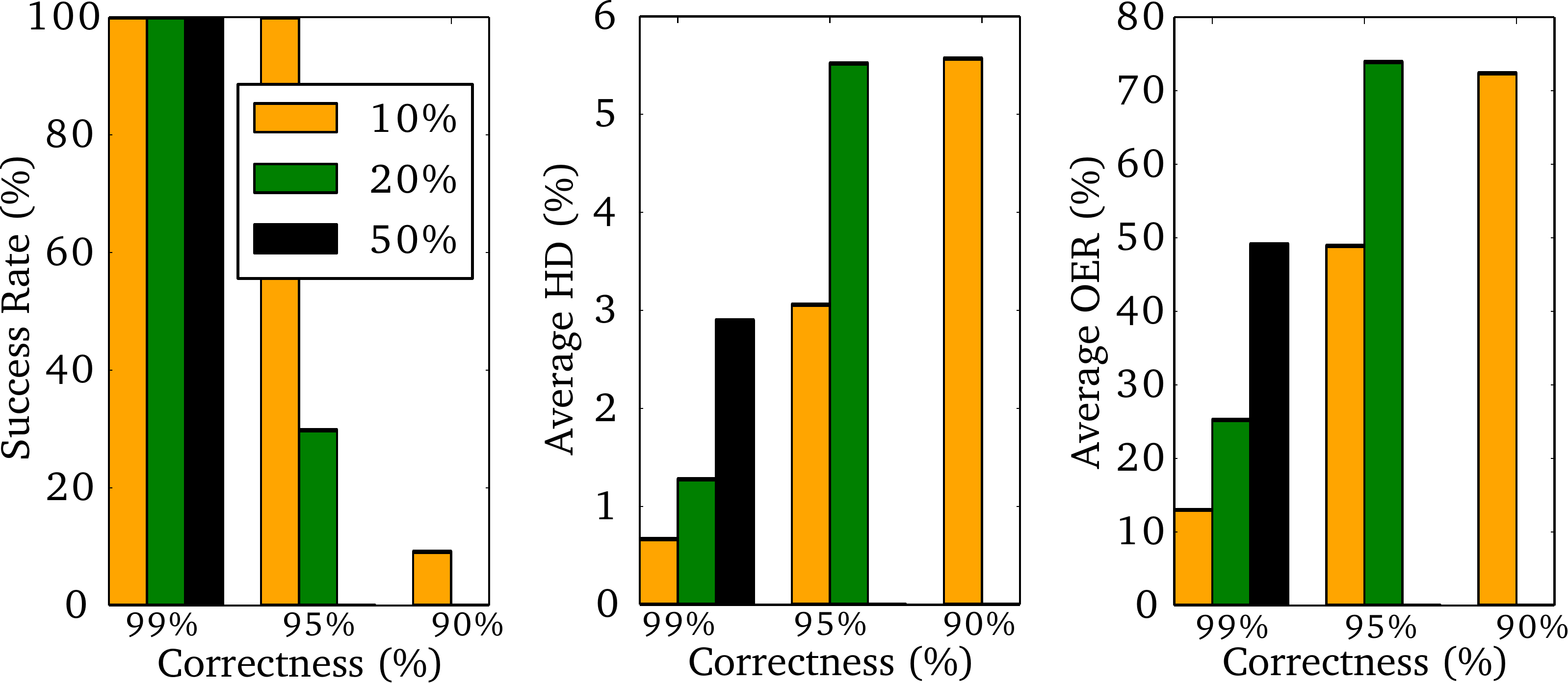}
}
}
\caption{Success rate, average HD, and average OER for 10,000 runs of our PSAT 
	attack~\cite{webinterface} applied on probabilistic versions of the \emph{ISCAS-85} benchmarks.
		See also Fig.~\ref{fig:conv_attack_prob_circuit} for the setup and metrics.
\label{fig:PSAT_prob_circuit}
}
\smallerspacecaption
\end{figure*}

\subsection{PSAT: Our Probabilistic SAT Attack}
\label{sec:PSAT}

Having demonstrated the ineffectiveness of the conventional SAT framework against probabilistic circuits, here we present our
advanced attack, called PSAT.
As indicated above,
   when tackling probabilistic circuits,
the main challenge for any SAT attack is to correctly prune the search space despite potential errors in the oracle's output.
An attacker cannot know in advance which output patterns are erroneous (since the circuit is obfuscated and, hence, initially of unknown
functionality). However, she/he can
apply each input pattern multiple times, track the statistical distribution of the output patterns, and pick only the most prevailing outputs
as \emph{ground truths} while pruning the search space during the SAT attack.
The basis for this assumption is that, in any probabilistic circuit, the overall error rate should be constrained such that the correct
output patterns occur more frequently than incorrect ones.
Otherwise, the circuit would become too approximate and might behave even arbitrarily at some point.

Accordingly, the fundamental concept of PSAT is \emph{Monte Carlo sampling}.
When querying the (probabilistic) oracle during the incremental SAT attack flow, we
apply each input pattern 1,000 times, without loss of generality, and we
track all resulting outputs.
Once this sampling is done, we sort the various output patterns by their number of occurrence. In case a
\emph{dominant pattern} is established,
we readily select this pattern as ground truth and proceed with the SAT attack.
An output pattern is considered dominant if and only if it occurs at least as many times as the second and third most frequent
patterns combined.
Otherwise, in case no dominant pattern is observed,
we randomly select among all observed patterns, but while considering their 
occurrence statistics.
For example, if an output pattern has been observed for 27\% of all oracle queries using the same input, this particular pattern has a 27\% random
chance to be selected as ground truth.

\subsubsection{\textbf{Experimental Setup}}

Similar as in Sec.~\ref{sec:SAT_attack_fail}, we implement PSAT
as an extension for the open-source framework of~\cite{subramanyan15,code_pramod}, and also here
we release our PSAT extension to the community as open source in return~\cite{webinterface}.
Along with the core techniques of PSAT, we provide supplementary features such as the
conversion of regular gates to probabilistic/polymorphic ones in the \emph{.bench} file format,
simulation of probabilistic/polymorphic gates, the sampling of Hamming distances, etc.
To enable a fair comparison between the conventional SAT attack and PSAT, we use the same probabilistic benchmark versions as
in Sec.~\ref{sec:SAT_attack_fail}, and we execute PSAT likewise for 10,000 times.

\subsubsection{\textbf{Results}}

As can be seen from Fig.~\ref{fig:PSAT_prob_circuit}, PSAT is notably more successful in resolving the obfuscation of the given
probabilistic circuits than the conventional attack.
For example, for 50\% of all gates being probabilistic ones with an error rate of 1\%, the conventional attack succeeded only for
4.3\% and 0.5\% of the 10,000 attack runs on \emph{c432} and \emph{c880} respectively. In contrast, PSAT holds a success rate of 100\% for those cases.
PSAT can resolve the underlying obfuscation with fewer errors, that is, the inferred key is more
accurate, and the behavior of the recovered IP matches matches more closely the original IP.
This is particularly the case for larger ranges of obfuscation using probabilistic gates. For example, for
50\% of all gates being probabilistic ones with an error rate of 1\%, the circuit \emph{c432} as recovered by the conventional attack has an HD and OER
of 11.5\% and 33.1\%, whereas the same circuit as recovered by PSAT has an improved HD and OER of 7.5\% and 23.0\%,
respectively. However, also PSAT it is challenged once the error rate of the probabilistic
gates increases to 10\%. That is because, for such a large error rate,
it is difficult to establish dominant patterns from the probabilistic oracle.
We observe similar results for \emph{apex4} and \emph{des}, which are not illustrated here due to lack of space.

These results reinforce our earlier argument on the trade-off between accuracy, power, and security against SAT attacks.
It is important to note that with excessive errors (of 10\% or more), the computational accuracy is rather low, limiting
the practicability of such overly imprecise circuits to begin with.

\begin{table}[tb]
\centering
\scriptsize
\setlength{\tabcolsep}{0.7mm}
\caption{Average Runtimes for Conventional SAT~\cite{code_pramod,webinterface} and PSAT}
\smallerspacecaption
\begin{tabular}{c|c|c|c|c}
\hline
\textbf{Benchmark} 
& \textbf{Camouflaging (\%)} 
& \textbf{Accuracy (\%)} 
& \textbf{Conventional SAT}
& \textbf{PSAT} \\ 
\hline \hline
c432 & 10 & 99  & 0.8752 s & 1.8153 s \\ \hline
c432 & 20 & 95 & 0.0411  s & 0.3525 s \\ \hline
c432 & 50 & 90 & 0.0296  s & 0.086  s \\ \hline
c880  & 10 & 99 & 1.0379 s & 2.8912 s \\ \hline
c880  & 20 & 95 & 0.0406 s & 1.4853 s \\ \hline
c880  & 50 & 90 & 0.0282 s & 0.0733 s \\ \hline
apex4 & 10 & 99 & 0.2044 s & 9.3149 s \\ \hline
apex4 & 20 & 95 & 0.1381 s & 0.6618 s \\ \hline
des   & 10 & 99 & 0.2374 s & 7.154  s \\ \hline
des   & 20 & 95 & 0.2675 s & 0.8273 s \\ \hline
\end{tabular}
\label{tab:runtime_numbers}
\end{table}

\
\subsubsection{Runtime}
\label{sec:PSAT_runtime}
Table~\ref{tab:runtime_numbers} shows the average runtime incurred by conventional SAT~\cite{code_pramod,webinterface} and PSAT, for few
	selected configurations, and across 10,000 runs.
As expected,
there are some runtime overheads due to Monte Carlo sampling, but also note the following.
First, the user/attacker is free to select less/more than 1,000 sampling runs; runtime overheads are controllable.
Second, the average runtime overheads are about 11.6X (across all configurations).
Third, the reported runtimes account already for the
final sampling runs (using 10,000 patterns, without loss of generality),
for HD and OER evaluation, which is not required for the conventional SAT attack.
	Overall, the runtimes and computational cost of PSAT can be considered affordable.

\subsection{PSAT on Polymorphic Circuits}
\label{sec:PSAT_counter}

Consider an embedded, reconfigurable design with dynamic time-sharing circuitry.  That is,
	 a single circuit is
tailored to perform all required operations serially on a time-sharing basis.
Such ``template circuitry'' can be readily implemented when leveraging the runtime polymorphism of GSHE-based circuits.
Moreover, operating the GSHE gates in the probabilistic regime would reduce the power consumption as well, rendering such circuitry a viable
scheme for low-power and error-tolerant Internet-of-Things (IoT) applications~\cite{blaauw2014iot}.

For such a scenario, even an advanced attack like PSAT may become ineffective.
That is because of the underlying principle of Monte Carlo sampling, which renders PSAT relatively slow,
depending on how many times each input pattern is to be evaluated and how fast the oracle can be physically queried.
Hence, for any iteration of the PSAT attack, the reconfigurable GSHE circuitry might have already morphed from
one logic structure to another, resulting in
an inconsistent oracle behavior. In turn, this is likely
to induce unsatisfiable assignments for the SAT model, 
causing PSAT to fail.

\subsubsection{\textbf{Experimental Setup}}
\label{sec:setup_polymorphic}

Prior setups are used as a baseline here.
We generate polymorphic versions of the benchmark circuits \emph{c432} and \emph{c880} as follows.
First, to provide a fair baseline, we re-apply the same random selection of gates as when picking 10\% for probabilistic gates in
Sec.~\ref{sec:SAT_attack_fail} and Sec.~\ref{sec:PSAT}.
Next, we configure each of those gates as polymorphic GSHE gates, while also accounting for their true functionality, to avoid excessive
overall errors.  That is, any NAND, AND, NOR, OR, XOR, or XNOR gate is replaced by a polymorphic NAND/AND/NOR/OR/XOR/XNOR gate, whereas the
original functionality has a probability of 66.67\% assigned, and all other functions a probability of 6.67\%. Likewise, any INV or BUF is
replaced by a polymorphic INV/BUF, whereas the original functionality has a probability of 66.67\% assigned as well.  We extend our
PSAT framework~\cite{webinterface} to account for such polymorphic behavior.
Using these baseline benchmarks containing 10\% polymorphic gates, we derive further benchmarks with 9\% down to 1\% polymorphic gates, in
steps of 1\%. We do so by randomly selecting polymorphic gates and reverting them to regular gates.
Like in the prior setups, we execute PSAT 10,000 times for each benchmark.

\subsubsection{\textbf{Results}}

\begin{figure}[tb]
\centering
\includegraphics[width=.85\textwidth]{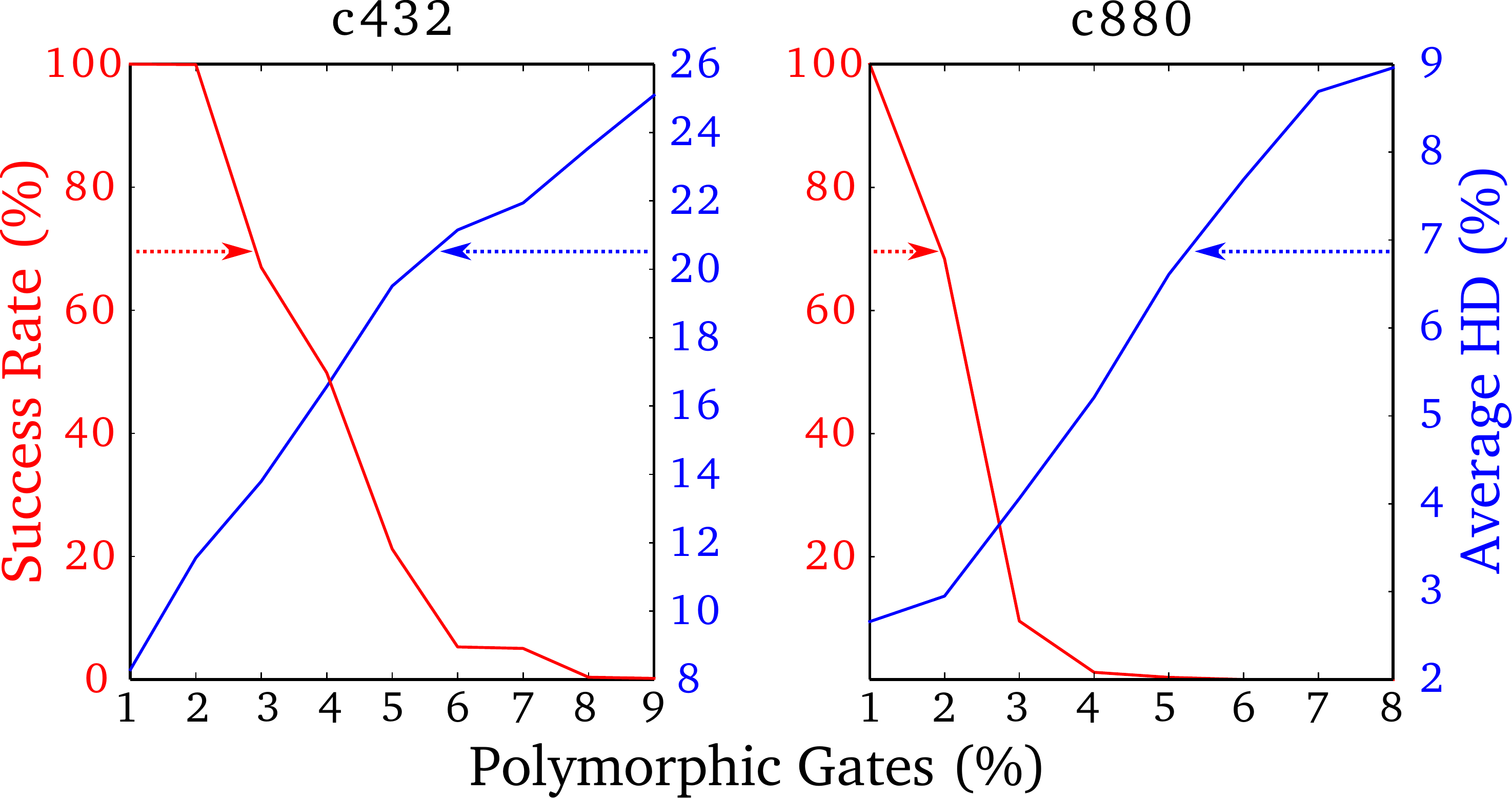}
\caption{Success rate (left axis, red) versus Hamming distance (HD, right axis, blue) for PSAT tackling circuits embedded with some polymorphic gates.
\label{fig:polymorphic_PSAT}
}
\smallerspacecaption
\end{figure}

As one would expect, PSAT can successfully resolve smaller scales of polymorphic obfuscation (Fig.~\ref{fig:polymorphic_PSAT}).
For larger scales, however,
the success rate is limited.
For those cases where PSAT is
successful, we also observe larger HD values than for the probabilistic scenario in Sec.~\ref{sec:PSAT}.
That is because polymorphic gates
tend to induce larger overall errors.

Similar to the probabilistic scenario, there is a trade-off between accuracy, power, and attack resilience.
While these experiments have shown a strong resilience against PSAT, we caution that
once polymorphic gates were used in a more predictable manner (as for example envisioned above for some embedded and reconfigurable design),
an attacker could tailor PSAT accordingly.
Thus, once the designer seeks to employ polymorphic gates while keeping the error in bounds, there may be further
protection measures required.

\section{Discussion}
\label{sec:discussion}

\subsection{Error Control for Probabilistic Circuits}
\label{sec:error_control}

\begin{figure}[tb]
\centering
	\includegraphics[scale=0.135]{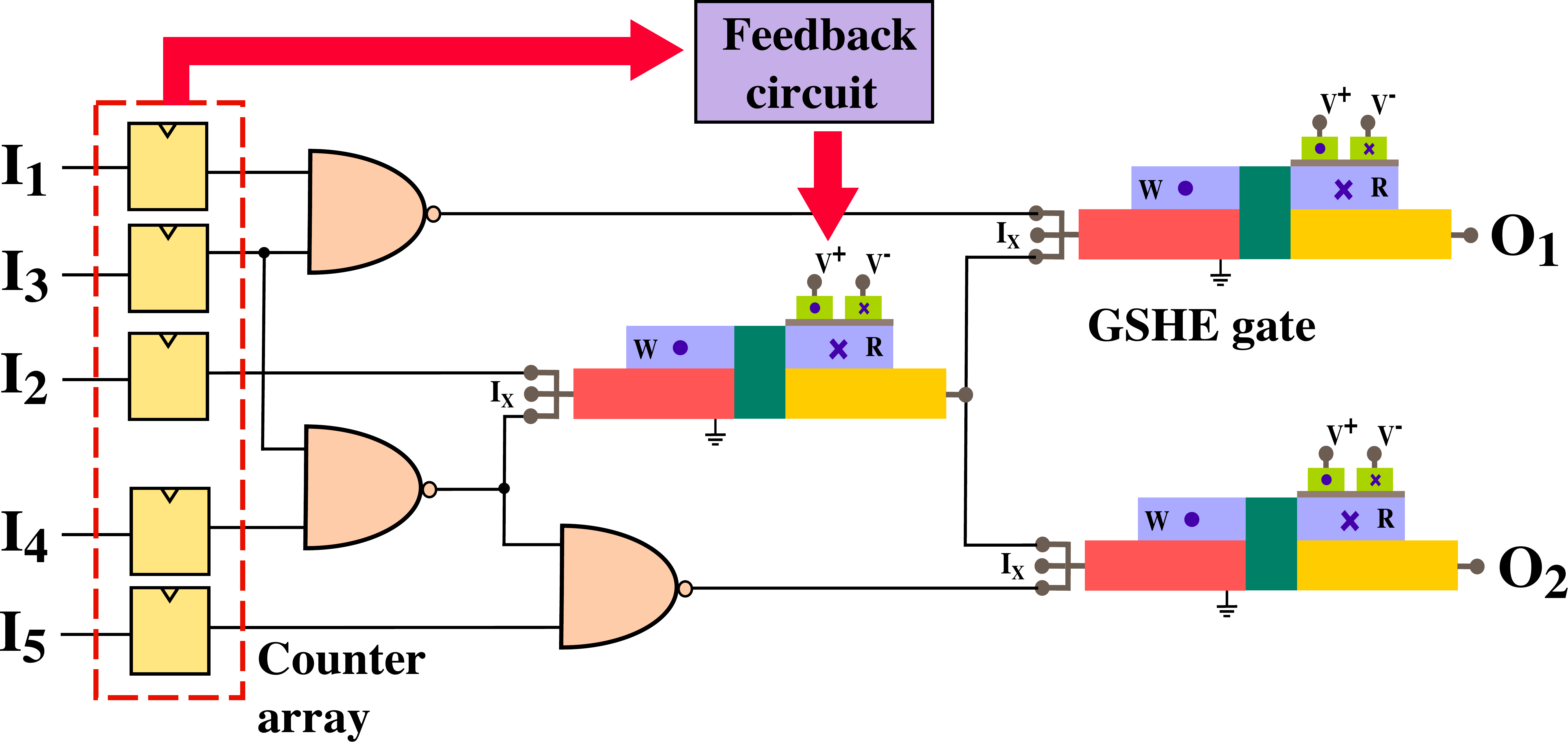}
\caption{
	Concept for a counter-based security mechanism.
	The input lines hold a counter to monitor suspicious pattern insertion.
	For clarity, the GSHE device has been laid out horizontally.
\label{fig:counter_defense}
}
\smallerspacecaption
\end{figure}

Here we envision a mechanism to detect repetitive sampling
	and to subsequently
scramble the statistical properties
of the GSHE circuit at runtime.
For example, consider the modified circuit \emph{c17} in Fig.~\ref{fig:counter_defense}, with three probabilistic NAND gates.
The error rates for the primary outputs $O_1$ and $O_2$ can be altered by tuning the MTJ voltages of the GSHE gate in the previous stage
(recall Sec.~\ref{sec:GSHE_operation}).
This feature is exploited in the conceptional feedback mechanism in Fig.~\ref{fig:counter_defense}, where a counter array placed at the
primary inputs checks the applied input patterns for overly repetitive patterns (or any other deviation from the expected,
application-specific inputs distribution, for that matter).
In case such suspicious behavior is observed, the feedback mechanism shall
dynamically adapt the MTJ voltage supply for the middle GSHE gate.
This would alter this gate's output current and, in turn, impact the error rate for the following two GSHE gates. Those latter two gates
then directly impact the overall circuit error rate, which can thwart advanced attacks such as PSAT.

The feedback mechanism itself should be implemented using deterministic but obfuscated GSHE gates which could help to obstruct removal
attacks. Moreover, the feedback mechanism should be integrated into the circuit in such a way that even advanced removal attacks
would result in fully stochastic circuit behavior. For example, the mechanism could be implemented such that removing it would also cut off the control current for the GSHE gates, which leads to unpredictable behavior of those gates.

\subsection{Other Advanced Attack Schemes}
\label{sec:other_attacks}

Besides our PSAT framework~\cite{webinterface}, we acknowledge that other attack schemes might be extended as well toward
resolving obfuscation in probabilistic and polymorphic circuits.
Although tailored to attack SAT-resilient schemes,
\emph{AppSAT}, which was recently proposed by Shamsi \emph{et al.}~\cite{shamsi18_TIFS}, is of particular interest.
That is because AppSAT is based on the probably-approximately-correct (PAC) paradigm, which can tolerate some
errors.\footnote{Shamsi \emph{et al.} tackle so-called compound schemes, e.g., \cite{yasin16_SARLock, xie16_SAT}.
	These schemes combine regular, large-error obfuscation techniques with provably secure, low-error obfuscation techniques.
		Shamsi \emph{et al.} have shown that PAC can help to reduce these schemes to their low-error obfuscation component.}
However, the attack is still relying on a consistent oracle behavior---an assumption which probabilistic circuits
do not adhere to. 
At the time of writing, the source code of AppSAT has not been available to us; hence, we were unable to incorporate
modeling of probabilistic and polymorphic circuit behavior into AppSAT.

More broadly, machine learning is increasingly being used for both developing new attacks
(e.g.,~\cite{lerman2011side,ganji17_thesis,zhang18,chakraborty2018sail}) and to
defend against attacks (e.g.,~\cite{sabhnani2003application}).
Whether machine learning attacks will be sufficiently robust and capable against probabilistic and polymorphic obfuscation schemes, however,
	remains to be seen.
For our scheme, we would like to point out that
(i)~the GSHE device experiences thermally induced stochasticity, i.e., truly random behavior~\cite{rangarajan2017energy},
(ii)~the error rate for any device can be tuned individually, and
(iii)~those individual error distributions superpose with each other while they propagate throughout the entire circuit.

\subsection{Case Study for Securing Hybrid GSHE-CMOS Circuits}
\label{sec:GSHE-CMOS}

Finally, we outline the prospects for securing industrial circuits using a hybrid GSHE-CMOS approach.
While the manufacturing of spin devices is still in nascent stages~\cite{matsunaga08,baek18,fong16}, such a
hybrid approach appears practical, given the CMOS-compatible processing of spin devices~\cite{matsunaga08,makarov16,parveen17}.

\begin{figure}[tb]
\centering
\includegraphics[width=.90\textwidth]{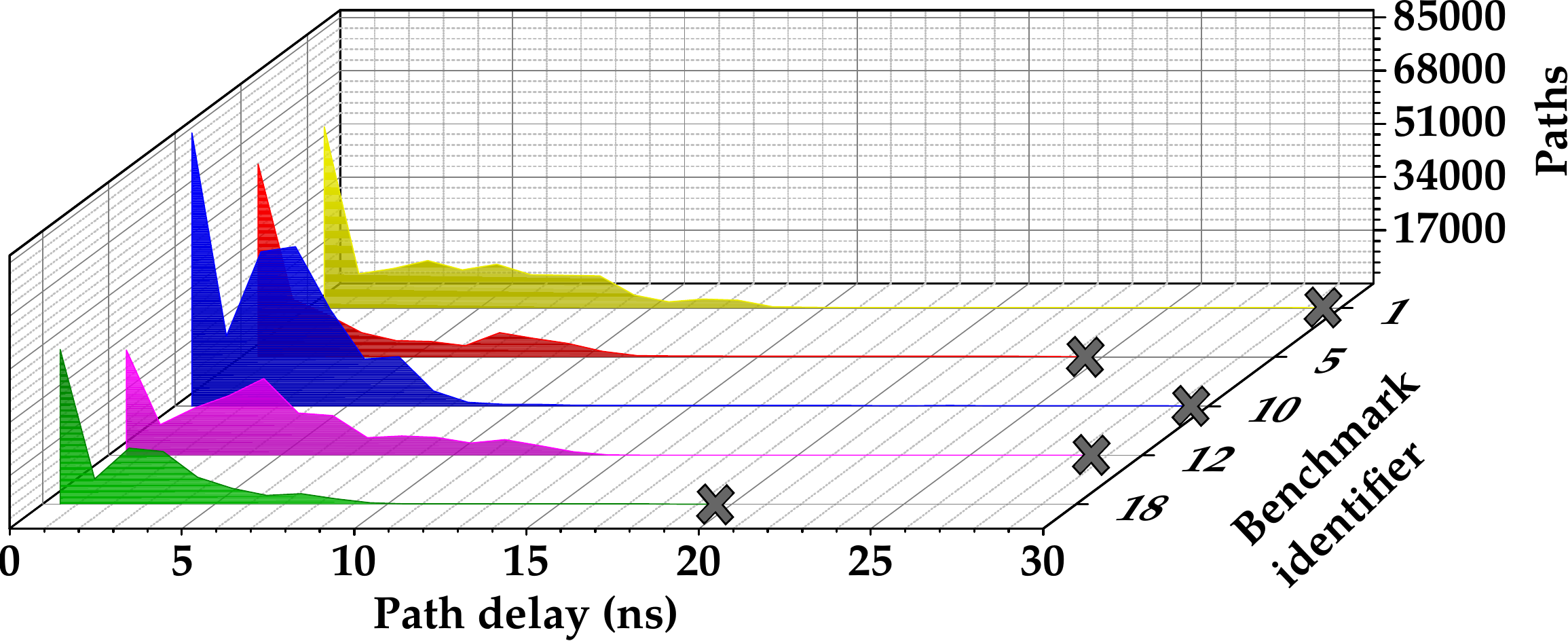}
\caption{Delay distributions of selected \emph{IBM superblue} circuits.
	For clarity, the paths with the critical delays are marked with crosses.
\label{fig:delay_superblue}
}
\smallerspacecaption
\end{figure}

On the one hand, recall that the delay for the GSHE device is larger than for regular CMOS devices (Sec.~\ref{sec:characterization}).
On the other hand, note that large industrial circuits tend to exhibit a skewed distribution of timing paths, with most paths imposing short
delays, and only a few paths inducing critical delays (Fig.~\ref{fig:delay_superblue}).
Here we explore a delay-aware approach for protecting the industrial \emph{IBM superblue} circuits~\cite{viswanathan11}.\footnote{
To process the layouts for the \emph{IBM superblue} circuits,
we leverage scripts provided by Kahng \emph{et al.}~\cite{kahng14}.
Being sequential circuits, we also have to pre-process them (to mimic access to the scan chains for the SAT attacks):
the inputs (and outputs) of any flip-flop are transformed into pseudo-primary outputs (and inputs), whereupon the flip-flops can be removed.
}
In short, 
we replace CMOS gates in the non-critical paths with an obfuscated, deterministic GSHE primitive,
as long as no delay overheads are incurred.
Doing so, we can obfuscate on average 5--15\% of all the gates in the benchmarks.

Conducting the conventional SAT attacks~\cite{subramanyan15,code_pramod} on those GSHE-augmented (but fully deterministic) designs, we
observe that they cannot be resolved within 240 hours. In fact, most attack trials incur computational limitations as previously indicated
in Sec.~\ref{sec:det_SAT_study}.
Although we select the \emph{IBM superblue} circuits here to showcase our delay-aware GSHE-CMOS hybrid camouflaging technique, this
approach can be applied to secure any large-scale circuit at little delay cost.

\section{Conclusion}
\label{sec:conclusion}

Imprecise computing is rapidly gaining traction
due to its attractive low-power characteristics.
In this paper, 
we present the first study exploring the hardware security prospects
of imprecise computing systems, specifically for probabilistic and polymorphic circuits constructed with noisy GSHE gates.
We design a GSHE-based primitive for IP protection by means of layout obfuscation. We provide not only a thorough security
analysis for this primitive,
mainly using conventional SAT attacks and our advanced attack PSAT, but we also discuss the inherent resilience of the GSHE device against
side channel attacks.
A key finding of this study is the following trade-off: the lower the accuracy of imprecise gates, the lower their power consumption, and
the better their resilience. Since any design may have practical limitations on the error tolerance, we also promote large-scale obfuscation
in the deterministic regime.
That is underpinned by another key finding of this study, namely that large-scale obfuscation can be competitive against provably
secure schemes.
Overall, we demonstrate imprecise computing, magnetic devices, and large-scale obfuscation as promising candidates for security-aware design requirements.

\section*{Acknowledgements}
This work was supported in part by the Semiconductor Research Corporation (SRC) and the National Science Foundation (NSF) through ECCS
1740136.  This work was carried out in part on the High Performance Computing (HPC) resources at New York University Abu Dhabi (NYUAD), and the
authors also acknowledge support from the Center for Cyber Security Abu Dhabi (CCS-AD) at NYUAD.

%\def\bibfont{\footnotesize}
%\bibliography{main,main2}
%\bibliographystyle{IEEEtran}

\begin{IEEEbiography}[{\includegraphics[width=1in,height=1.25in,clip,keepaspectratio]{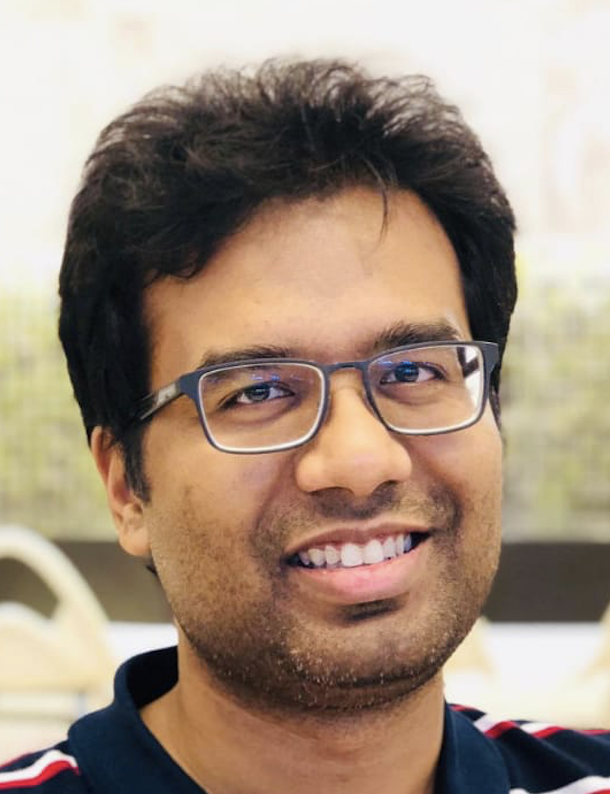}}]{Satwik Patnaik} (M'16)
received B.E.\
in Electronics and Telecommunications from the University of
Pune, Pune, India and
M.Tech.\ in Computer Science and Engineering with a specialization in VLSI Design from Indian Institute of Information Technology and
Management, Gwalior, India. 
He is a Ph.D.\ candidate at the Department of Electrical and Computer Engineering at the 
Tandon School of Engineering with New York University, Brooklyn, 
NY, USA. 
He is a Global Ph.D.\ Fellow with New York University Abu Dhabi, Abu Dhabi, UAE. 
His current research interests 
include Hardware Security, Trust and Reliability issues for CMOS and Emerging Devices 
with particular focus on low-power VLSI Design.
He is a student member of IEEE and ACM.
\end{IEEEbiography}

\begin{IEEEbiography}[{\includegraphics[width=1in,height=1.25in,clip,keepaspectratio]{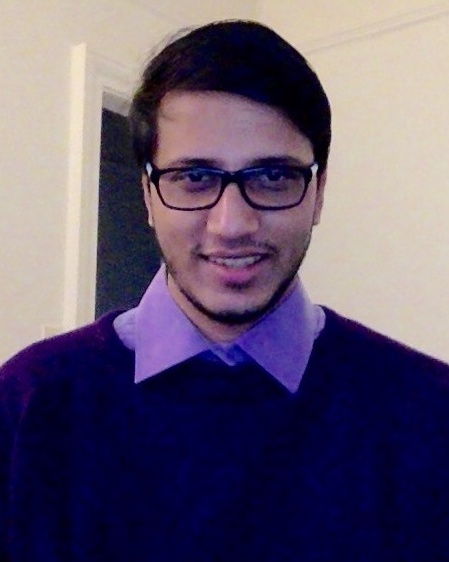}}]{Nikhil Rangarajan}
is a Ph.D.\ candidate at the department of Electrical and Computer engineering, New York University, New York, NY, USA. 
He has an M.S.\ in electrical engineering from New York University. 
His research interests include spintronics, nanoelectronics and device physics. 
He is a student member of IEEE. %and he can be reached at nikhil.rangarajan@nyu.edu. 
\end{IEEEbiography}

\begin{IEEEbiography}[{\includegraphics[width=1in,height=1.25in,clip,keepaspectratio]{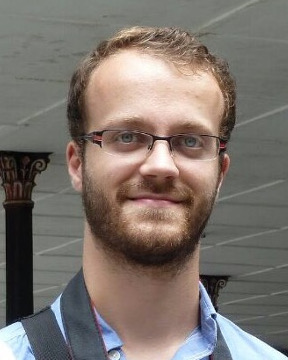}}]{Johann Knechtel}
(M'11)
received the M.Sc.\ in Information Systems Engineering (Dipl.-Ing.) in 2010 and the Ph.D.\ in Computer Engineering
(Dr.-Ing.) in 2014, both from TU Dresden, Germany.  He is currently a Research Associate
at the New York University Abu Dhabi (NYUAD), UAE.  Dr.\ Knechtel was a
Postdoctoral Researcher in 2015--16 at the Masdar Institute of Science and Technology, Abu Dhabi.  From 2010 to 2014, he was
a Ph.D.\ Scholar with the DFG Graduate School on ``Nano- and Biotechnologies for Packaging of Electronic
Systems'' and the Institute of Electromechanical and Electronic Design, both hosted at the TU Dresden.  In 2012, he was a
Research Assistant with the Dept.\ of Computer Science and Engineering, Chinese University of Hong Kong, China.  In 2010, he
was a Visiting Research Student with the Dept.\ of Electrical Engineering and Computer Science, University of Michigan, USA.
His research interests cover VLSI Physical Design Automation, with particular focus on Emerging Technologies and Hardware Security.
\end{IEEEbiography}

\begin{IEEEbiography}[{\includegraphics[width=1in,height=1.25in,clip,keepaspectratio]{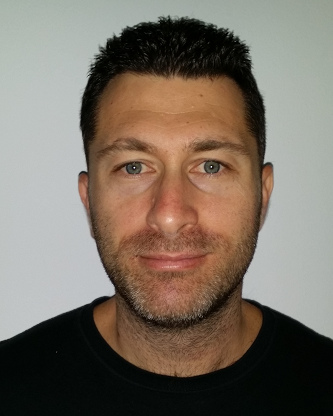}}]{Ozgur Sinanoglu}
is a Professor of Electrical and Computer Engineering at New York University Abu Dhabi. He earned his B.S.\ degrees, one in
Electrical and Electronics Engineering and one in Computer Engineering, both from Bogazici University, Turkey in 1999. He obtained his MS
and PhD in Computer Science and Engineering from University of California San Diego in 2001 and 2004, respectively. He has industry
experience at TI, IBM and Qualcomm, and has been with NYU Abu Dhabi since 2010. During his PhD, he won the IBM PhD fellowship award twice.
He is also the recipient of the best paper awards at IEEE VLSI Test Symposium 2011 and ACM Conference on Computer and Communication Security
2013. 

Prof.\ Sinanoglu's research interests include design-for-test, design-for-security and design-for-trust for VLSI circuits, where he has more
than 180 conference and journal papers, and 20 issued and pending US Patents. Sinanoglu has given more than a dozen tutorials on hardware
security and trust in leading CAD and test conferences, such as DAC, DATE, ITC, VTS, ETS, ICCD, ISQED, etc. He is serving as track/topic
chair or technical program committee member in about 15 conferences, and as (guest) associate editor for IEEE TIFS, IEEE TCAD, ACM JETC,
      IEEE TETC, Elsevier MEJ, JETTA, and IET CDT journals. 

Prof.\ Sinanoglu is the director of the Design-for-Excellence Lab at NYU Abu Dhabi. His recent research in hardware security and trust
is being funded by US National Science Foundation, US Department of Defense, Semiconductor Research Corporation, Intel Corp and
Mubadala Technology.
\end{IEEEbiography}
\pagebreak

\begin{IEEEbiography}[{\includegraphics[width=1in,height=1.25in,clip,keepaspectratio]{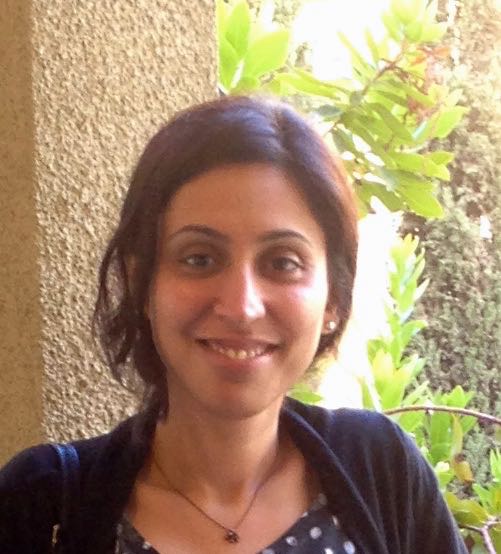}}]{Shaloo Rakheja}
is an Assistant Professor of Electrical and Computer engineering with New York University, Brooklyn, NY, USA, where she works on
nanoelectronic devices and circuits. Prior to joining NYU, she was a Postdoctoral Research Associate with the Microsystems Technology
Laboratories, Massachusetts Institute of Technology, Cambridge, USA. She obtained her M.S.\ and Ph.D.\ degrees in Electrical and Computer
Engineering from Georgia Institute of Technology, Atlanta, GA, USA. 
\end{IEEEbiography}

\end{document}